\documentclass[12pt]{article}
\pdfoutput=1
\usepackage[top=1in, bottom=1in, left=1.in, right=1.in]{geometry}
\usepackage[english]{babel}
\usepackage{atbegshi,cite}
\usepackage{amsmath,amssymb,amsbsy,amstext, amsthm, simplewick}
\usepackage{hyperref}
\usepackage{graphicx}
\usepackage{amsfonts}
\usepackage[small]{caption}
\usepackage{upgreek}
\usepackage[titletoc]{appendix}
\usepackage{setspace}
\usepackage{color}
\usepackage{slashed}

\usepackage{caption}
\usepackage{subcaption}

\theoremstyle{plain}
\theoremstyle{plain} 
\theoremstyle{plain} 
\theoremstyle{plain}
\theoremstyle{plain}
\theoremstyle{plain}

\renewcommand{\title}[1]{{\Large\bf\flushleft{#1}}\vspace*{3ex}\\}
\renewcommand{\author}[2]{{\noindent\hspace*{2.5em}\large#1}
                     \footnote{Electronic mail: $\mathtt{#2}$}\\}

\newcommand{\newc}{\newcommand}
\newc{\fpi}{f_{\pi}}

\newc{\etap}{\eta^{\prime}}
\newc{\llll}{\langle\lambda\lambda\rangle}
\newc{\FFd}{F^a\tilde F^a}
\newc{\qbar}{{\overline q}}
\newc{\TR}{{\rm Tr}}
\newc{\Kahler}{K\"ahler }
\newc{\Zbb}{{\mathbb Z}}
\newc{\Rt}{{{\mathbb R}^3}}
\newc{\Rf}{{{\mathbb R}^4}}
\newc{\Sth}{{{\mathbb S}^3}}
\newc{\Sn}{{{\mathbb S}^n}}
\newc{\SthSo}{{{\mathbb S}^3\times{\mathbb S}^1}}
\newc{\Stw}{{{\mathbb S}^2}}
\newc{\StwSo}{{{\mathbb S}^2\times{\mathbb S}^1}}
\newc{\So}{{{\mathbb S}^1}}
\newc{\zt}{{{\mathbb Z}_2}}
\newc{\RtSo}{{{\mathbb R}^3\times{\mathbb S}^1}}
\newc{\RfSo}{{{\mathbb R}^4\times{\mathbb S}^1}}
\newc{\RfSn}{{{\mathbb R}^4\times{\mathbb S}^n}}
\newc{\scriminus}{{\cal I}^-}
\newc{\scriplus}{{\cal I}^+}
\newc{\mpl}{M_p}
\newc{\Ricci}{\mathcal{R}}
\newc{\bv}{\phi}
\newc{\calU}{{\cal U}}
\newc{\calK}{K}
\newc{\calUi}{{\cal U}^{-1}}
\newc{\calG}{{\cal G}}
\newc{\calI}{{\cal I}}
\newc{\calO}{{\cal O}}
\newc{\calM}{{\cal M}}
\newc{\calQ}{{\cal Q}}
\newc{\calT}{{\cal T}}
\newc{\calOb}{{\cal O}^\dagger}
\newc{\hphi}{{\hat\phi}}

\newcommand{\xib}{{\xi_b}}
\newcommand{\beq}{\begin{equation}}
\newcommand{\eeq}{\end{equation}}

\begin{document}
\begin{titlepage}
\begin{flushright}
{\large 
~\\
}
\end{flushright}

\vskip 2.2cm

\begin{center}

{\large \bf Neutralizing Topological Obstructions to Bubbles of Nothing}

\vskip 1.4cm

{{Patrick Draper}$^1$, {Benjamin Lillard}$^{1,2}$, and {Carissa Skye}$^1$}
\\
\vskip 1cm
$^1$\textit{Department of Physics, University of Illinois, Urbana, IL 61801}
\vskip 4pt
$^2$\textit{Institute for Fundamental Science and Department of Physics,\\ University of Oregon, Eugene, OR 97401}
\vspace{0.3cm}
\vskip 4pt

\today

\vskip 1.5cm

\begin{abstract}

Theories with compact extra dimensions can exhibit a vacuum instability known as a bubble of nothing. These decay modes can be obstructed if the internal manifold is stabilized by fluxes, or if it carries Wilson lines for background gauge fields, or if the instanton is incompatible with the spin structure. In each of these cases the decay can proceed by adding dynamical charged membranes or gauge fields. We give a general, bottom-up procedure for constructing approximate bubble of nothing solutions in models with internal spheres stabilized by flux and study the influence of the brane tension on the tunneling exponent, finding two branches of solutions that merge at a minimal superextremal value of the tension. In the case of Wilson operators and incompatible fermions, the relevant bubble is shown to be the Euclidean Reissner-Nordstrom black hole, and the ordinary decay exponent is modified by $1/g^2$ effects. We examine the Dirac operator on this background and comment on the relevance for models of supergravity with gauged $R$-symmetry.

\end{abstract}

\end{center}

\vskip 1.0 cm

\end{titlepage}
\setcounter{footnote}{0} 
\setcounter{page}{1}
\setcounter{section}{0} \setcounter{subsection}{0}
\setcounter{subsubsection}{0}
\setcounter{figure}{0}



\section{Introduction}

The gravitational force  enriches the semiclassical theory of vacuum decay~\cite{Coleman:1980aw}. An interesting class of decay processes in gravitational theories involves changes in the topology of spacetime. A famous example arises in the pure Kaluza-Klein (KK) theory, which admits an instability called a bubble of nothing (BON)~\cite{Witten:1981gj}. The gravitational instanton is equivalent to the 5D Euclidean Schwarzschild geometry, and it represents a topology-changing transition that puts a hole in spacetime. Similar bubble of nothing-type instantons exist in a wide range of extra-dimensional settings with a variety of internal manifolds~\cite{Fabinger:2000jd,Brill:1991qe,DeAlwis:2002kp,Horowitz:2007pr,BlancoPillado:2010df,BlancoPillado:2010et,Brown:2010mf,Blanco-Pillado:2016xvf,Ooguri:2017njy,Acharya:2019mcu, GarciaEtxebarria:2020xsr,Bomans:2021ara}.

BON transitions can be obstructed by topological effects associated with gauge and matter fields. If the geometric moduli of the internal manifold are stabilized by flux, the quantization $\int F = 2\pi \mathbb{Z}$ prevents them from smoothly capping off in the bubble geometry. A background Wilson operator $\int A$ around the compact dimensions can also forbid them from smoothly vanishing. Closely related to the latter, it was pointed out already in~\cite{Witten:1981gj} that fermion spin structures can also be incompatible with some BON decays.

In each of these cases, couplings to other dynamical fields permit the BON vacuum decay to proceed despite the topological obstruction from flux or incompatible fermionic spin structures. In this work we use bottom-up techniques and simple examples to study these modified bubble nucleation processes and their leading tunneling exponents.

In the first part of our work we study flux-stabilized potentials.  
If suitable charged membrane sources are present in the theory, hybrid membrane-BON nucleation processes can proceed, simultaneously screening the flux and changing the topology. Examples of such bubbles were studied in~\cite{BlancoPillado:2010et,Brown:2010mf} in the context of 6D Einstein-Maxwell theory, and in~\cite{Bomans:2021ara} for the compactification of 10d supergravity onto the $S^6$ sphere.
Here we develop a bottom-up procedure to construct bubble solutions for arbitrary internal $n$-spheres with arbitrary moduli potentials, extending some of the techniques in~\cite{Dine:2004uw,Draper:2021ujg,Draper:2021qtc} to problems with flux and membranes. The framework we develop is analytic, but generally the equations of motion and jump conditions must be solved numerically.
We carry out this procedure in some simple examples. An interesting new finding of the numerical analysis is the existence of multiple branches of BON solutions, and a novel \emph{lower} bound on the tension of the charged membrane, below which the vacuum is not destabilized by this Schwinger--BON decay. Above this minimum, we demonstrate the dependence of the tunneling exponent on the brane tension. 

In the second part of the paper we study the connection between Wilson operators and fermions. A  Wilson operator for a flat background gauge potential can arise in the low energy description of a theory with massive fermions, encoding the fermion spin structure in the infrared gravitational EFT. With either a background Wilson operator or explicit fermions, there is an obstruction to bubbles of nothing which can be lifted by giving the gauge field dynamics/gauging fermion number. We show that the simplest example of this phenomenon arises already 5D KK theory, where the relevant bubble is given exactly by 5D Euclidean Reissner-Nordstrom. Notably, we find that this BON decay can proceed in a specific example of softly broken gauged 5d supergravity. 

The problems we study are of general academic interest in the field of vacuum decay in higher dimensional spacetimes. Furthermore, in the context of a hypothetical string landscape, identifying all the decay modes for vacua supporting fluxes or Wilson operators, and being able to calculate the rates parametrically, is necessary in order to find candidates for our universe.

In the final section we summarize our primary findings and comment on some directions for future work.

\section{Bubbles of Nothing through Flux-Stabilized Potentials}
To study bubble of nothing decays in the presence of a general modulus-stabilizing potential, it is convenient to reformulate the instanton as a solution to a Coleman-De~Luccia (CDL) problem with nonstandard boundary conditions. This approach was first introduced in~\cite{Dine:2004uw} in the absence of flux and the general boundary conditions were given in~\cite{Draper:2021qtc,Draper:2021ujg}.  In this section we set up the problem with flux. We begin by deriving some useful dimensional reduction formulas needed for the analysis, including careful treatment of the on-shell gravitational action and its boundary terms, and  describe the  fluxes and brane sources of interest.

\subsection{Dimensional Reduction}

We are interested in $4+n$-dimensional  Euclidean Einstein gravity and its dimensional reduction to four dimensions. The   action is
\begin{align}
	S = M^{2+n}\int_{\calM}&d^{4+n} x \sqrt{g_{4+n}} \left(-  \frac{1}{2} \mathcal{R}_{4+n} + \Lambda^\text{CC}_{4+n} \right)-M^{2+n}\int_{\partial\calM} d^{3+n} x \sqrt{h_{3+n}}\left(K-K_0\right) +{\rm (sources)}
	\label{eq:fourplusnaction}
\end{align}
where $M$ is the fundamental Planck mass. The scalar-tensor sector of the Einstein frame dimensional reduction is obtained by substituting
\begin{equation}
	ds_{4+n}^2 = e^{ - \sqrt{\frac{2n}{n+2}} \frac{\phi}{M_p}} d s_4^2 + e^{ 2\sqrt{\frac{2}{n(n+2)}} \frac{\phi}{M_p} } d s_n^2 .
\label{eq:Eframe}
\end{equation}
Here the 4d Planck scale is $M_p^2 = M^{2+n}V_{n}$, where $V_{n}$ is the volume of the internal $n$-manifold. We will mostly be interested in $n$-spheres, for which we write the volume as $S_n$. 
Substituting~(\ref{eq:Eframe}) into~(\ref{eq:fourplusnaction}) we obtain the reduced action for the 4d metric $g$ and volume modulus $\phi$,
\begin{align}
S =& \int d^4 x \sqrt{ g} \left(
		-\frac{M_p^2}{2} \mathcal{R}
		+ \frac{1}{2} g^{\mu\nu} \partial_\mu \phi \partial_\nu \phi
		- \frac{M_p}{2} \sqrt{\frac{2n}{n+2}} \Box \phi
		+ U(\phi) \right) \nonumber\\
	&+\frac{1}{2}M_p\sqrt{\frac{2n}{n+2}} \int_{\partial \mathcal{M}} d^3x
	\sqrt{h}n^\mu\partial_\mu \phi +  S_{\rm GHY} .
\label{eq:Sreduced}
\end{align}
Here $ S_{\rm GHY} $ is a regularized  GHY term on a 3D boundary, and both boundary terms in the second line of~(\ref{eq:Sreduced}) arise from the  reduction of the $4+n$-dimensional GHY term in~(\ref{eq:fourplusnaction}). The $U(\phi)$ term in~(\ref{eq:Sreduced}) is a modulus potential which we write as
\begin{align}
	U(\phi) = -\frac{M_p^2}{2} \mathcal{R}_n e^{- \sqrt{\frac{2n + 4}{n}} \frac{\phi}{M_p}} + M_p^2 \Lambda_{4+n} e^{- \sqrt{\frac{2n}{n+2}} \frac{\phi}{M_p}} +  c e^{ - 3 \sqrt{\frac{2n}{n+2}} \frac{\phi}{M_p}} + U_{other}.
	\label{eq:sourcesofmodulipotentials}
\end{align}
For an internal $n$-sphere, $ds_n^2\equiv R_n^2 d\Omega_n^2$ and  $\mathcal{R}_n = n(n-1)/R_n^2$. Both the curvature term and the cosmological constant term arise from the gravitational part of the $4+n$-dimensional  action.  $c$ is a constant of order $Q^2$ counting $n$-form flux wrapping the internal manifold,
\begin{align}
c &= \frac{Q^2}{2 g^2 S_n}, 
\label{eq:cg}
\end{align}
and $U_{other}$ denotes sources of moduli potentials other than higher dimensional cc, curvature, and flux.  The flux term and $U_{other}$ arise from the ``sources" term in~(\ref{eq:fourplusnaction}), which we will not need to specify in more detail.

\subsection{CDL ansatz}
The CDL  ansatz for the 4d metric is spherically symmetric,
\begin{equation}
ds_4^2 = d\xi^2+\rho(\xi)^2 d\Omega_3^2 ,
\label{eq:cdlmetric}
\end{equation}
where $\xi$ is a radial coordinate and $\rho$ sets the curvature radius of the transverse 3-sphere. In terms of $\rho$ and $\phi$, the 4d bulk  action (first line of Eq.~(\ref{eq:Sreduced})) takes the form
\begin{align}
S_{bulk}=2\pi^2 \int d\xi\, \left[\rho^3\left(\frac{1}{2}\phi'^2-\frac{M_p}{2}\sqrt{\frac{2n}{n+2}}\Box\phi+U\right)+3M_p^2(\rho^2\rho''+\rho\rho'^2-\rho)\right]
\label{eq:SbulkCDL}
\end{align}
where $\Box \phi = \phi''+3\frac{\rho'}{\rho}\phi'$. 
To  rewrite in the boundary term $S_{\rm GHY}$ in the CDL ansatz, it is convenient to change the radial coordinate to the area radius, $r\equiv \rho(\xi)$. Then a suitable reference metric for use in the $4+n$ dimensional GHY subtraction term in Eq.~(\ref{eq:fourplusnaction}) is
\begin{align}
	ds_{4+n}^2 = e^{ - \sqrt{\frac{2n}{n+2}} \frac{\phi(\hat r)}{M_p}} (g_{rr} dr^2+r^2 d\Omega_3^2) + e^{ 2\sqrt{\frac{2}{n(n+2)}} \frac{\phi(\hat r)}{M_p} } d s_n^2 .
\label{eq:refmetric}
\end{align}
Here the modulus is evaluated at a large fixed boundary radius $\hat r$, so that the exponential functions are constant. The reference metric~(\ref{eq:refmetric}) has the property  that the induced metric on $\hat r$ is the same in the bubble and reference geometries. The reduced GHY term integrated over the 3-sphere works out to be
\begin{align}
S_{\rm GHY}  = -6\pi^2\hat r^2M_p^2\left(\frac{1}{\sqrt{g_{rr}(\hat r)}}-e^{ - \frac{1}{2}\sqrt{\frac{2n}{n+2}} \frac{\phi(\hat r)}{M_p}}\right)\bigg|_{\hat r\rightarrow\infty}.
\label{sghypresimplify}
\end{align}
We can simplify it further by assuming that $\phi\rightarrow 0$ faster than $1/r^2$ at large radius, which will be the case for stabilized moduli, and that the CDL metric function $\rho \approx \xi + u + v/\xi$ at large radius for some constants $u$ and $v$, so that $\xi\approx r-u-v/r$ and $1/\sqrt{g_{rr}(\hat r)} \approx 1-v/\hat r^2 $. Then we find that Eq.~(\ref{sghypresimplify}) reduces to
\begin{align}
S_{\rm GHY}= 6\pi^2v^2M_p^2.
\label{eq:sghyformula}
\end{align}
This is different from Witten's bubble, where the modulus is not stabilized and the extrinsic curvature of the reference geometry cancels that of the bubble, so that $S_{\rm GHY}=0$. However in many case it turns out that $v=0$.

Putting the pieces together, we can write the total CDL action as
\begin{align}
S &= S_{bulk} + S_{\phi \scriptscriptstyle{bndry}} + S_{GHY}
\end{align}
where $S_{\phi \scriptscriptstyle{bndry}}$ is the first term in the second line of~(\ref{eq:Sreduced}). However, for constructing the action and deriving the equations of motion in the presence of a brane, it is also  convenient to rewrite the bulk action in a first-order form, integrating the bulk total derivative term $\Box\phi$ and integrating by parts the term proportional to $\rho''$. Ordinarily, writing the action in first-order form removes most of the surface terms, and integrating   $\Box\phi$ and $\rho''$ by parts does remove $S_{\phi{ \tiny bndry}} + S_{GHY}$.\footnote{More precisely, it removes the $K$ part of the GHY term; the $K_0$ subtraction term is retained and is important to cancel divergences in the on-shell action.} However, bubbles of nothing have fake singularities at $\xi=0$ in dimensionally reduced form, resulting in two residual  terms in the first-order form:
\begin{align}
S_{bulk}&\rightarrow  2\pi^2 \int_0^\infty d\xi\, \left[\rho^3\left(\frac{1}{2}\phi'^2+U\right)-3M_p^2(\rho\rho'^2+\rho)\right]\nonumber\\
&~~~~~~~~+6\pi^2M_p^2\rho^2\rho'\big|_{\xi=0}+\pi^2M_p\sqrt{\frac{2n}{n+2}} \rho^3\phi'\big|_{\xi=0}.
\label{eq:sbulkfirstorder}
\end{align}
This form is most useful for deriving the equations of motion. However,  when we compute the on-shell action, we will use an intermediate form of Eq.~\eqref{eq:SbulkCDL}, retaining the second-order term proportional to $\rho''$ but integrating out $\Box\phi$. Then only the final term in (\ref{eq:sbulkfirstorder}) appear at $\xi=0$, and the full $S_{GHY}$ is retained at $\xi\rightarrow\infty$.

\subsection{Flux compactifications and screening by branes}

We focus on the case where the internal $n$-sphere is supported by $n$-form flux,
\begin{align}
\int_n F = Q.
\end{align}
The gauge kinetic term is
\begin{align}
-\frac{1}{2g^2}\int_{4+n} F\wedge \star F. 
\end{align}
Integrating the kinetic term just over the $n$-sphere supplies a contribution to the $\phi$ potential. We have
\begin{align}
-\frac{1}{2g^2}\int_n F\wedge \star F =-\frac{Q^2}{2g^2 S_n}e^{-\sqrt{\frac{2n}{n+2}}\frac{\phi}{M_p}}
\label{gaugekinetictoUflux}
\end{align}
where $S_n e^{\sqrt{\frac{2n}{n+2}}\frac{\phi}{M_p}}$ is the sphere area, with $S_n=\frac{2\pi^{\frac{n+1}{2}}}{\Gamma\left(\frac{n+1}{2}\right)}(R_n)^n$.  
To obtain $U_{flux}$, we must combine (\ref{gaugekinetictoUflux}) with the $\phi$ dependence of the 4-dimensional submetric, the volume measure of which provides an additional factor of $e^{-2\sqrt{\frac{2n}{n+2}}\frac{\phi}{M_p}}$, and add a minus sign. 
We find
\begin{align}
U_{flux} = \frac{Q^2}{2g^2 S_n}e^{-3\sqrt{\frac{2n}{n+2}}\frac{\phi}{M_p}} ,
\end{align}
from which we can read off $c= \frac{Q^2}{2g^2 S_n}$ in the notation of Eq.~(\ref{eq:sourcesofmodulipotentials}). 

As a result, the modulus potential in the dimensionally-reduced formulation has a term that grows rapidly in the compactification limit $\phi\rightarrow-\infty$. The CDL equations have no smooth solution in the bubble-of-nothing class in the presence of this term~\cite{Draper:2021qtc,Draper:2021ujg}. However, a bubble can co-nucleate with a charged brane or a stack of charged branes that screens the flux entirely. In the case of $n$-form flux in $4+n$ dimensions we require magnetically charged $2$-branes.  In the Euclidean geometry we wrap the brane worldvolume over the $S^3$ bubble at some fixed radius $\xib$, to be determined dynamically.

The brane sits at a point on the internal $n$-sphere, breaking the internal rotational symmetries. This complicates the problem and leads to an imperfect cancellation of the flux energy  inside $\xib$. ($\int F=0$ by the Gauss law, but $\int F^2\neq 0$ because of the symmetry breaking.)  This  was also noted in~\cite{Brown:2010mf}. Unfortunately, to systematically improve the treatment would introduce enormous technical complication due to the loss of symmetry. Therefore, as in~\cite{Brown:2010mf}, we  work in the thin-brane limit and in an approximation where all of the flux is screened inside $\xib$. We comment briefly on extensions of this treatment in the conclusions.

For complete screening we require either a single brane of charge $Q$ or a stack of coincident branes of total charge $Q$.  
Due to brane-brane gravitational and electromagnetic interactions, the total tension $T$ of the stack is not 
generally the sum of the tensions of the individual branes. 
In the extremal limit these interactions cancel, however, and in any case our analysis depends only on the total tension. 
Furthermore our interest is in superextremal branes: if there are fundamental branes they will be less expensive to nucleate than black branes, and among the black branes the extremal one should provide the fastest rate~\cite{Brown:2010mf}. 
The general extremality bound is~\cite{Garfinkle:1990qj,Heidenreich:2015nta}
\begin{align}
\frac{ 8\pi G T^2}{Q^2/g^2} < \frac{1}{3}\left(\frac{n+2}{n-1}\right)
\end{align}
where $G$ is the $4+n$-dimensional Newton constant, $8\pi G = M^{-2-n}$. In terms of the potential coefficient $c$,
\begin{align}
\frac{Q^2/g^2}{8\pi G T^2} = \frac{2 S_n c}{8\pi G T^2}= \frac{2  c}{8\pi G_{(4)} T^2}.
\end{align}

\subsection{Equations of motion}

To construct instantons describing the co-nucleation of bubbles and branes, we use the CDL equations in the dimensionally reduced description. As described above we work in the simplifying approximation where the brane is thin and screens all of the flux inside its nucleation radius, $\xi=\xi_b$ in terms of the CDL radial coordinate. This means $\xi_b$ acts as a junction across which the potential and fields have mild discontinuities. We construct the equations of motion and jump conditions and then describe approximate solutions.

To obtain the equations of motion, it is convenient to vary the first-order form of the bulk action, because it does not incur Dirac delta singularities even in the presence of the  metric discontinuities induced by the thin brane. We write the bulk  action as an exterior part, $\xi>\xib$, where the flux is nonzero and the total potential is $U$; an interior part, $\xi<\xib$, where the flux is screened and we set $U=U|_{Q=0}$; and a brane worldvolume contribution at $\xib$. The complete bulk action is then:
\begin{align}
S_{bulk}&=2\pi^2 \int_0^\xib d\xi\, \left[\rho^3\left(\frac{1}{2}\phi'^2+U|_{Q=0}\right)-3M_p^2(\rho\rho'^2+\rho)\right]\nonumber\\
 &~~+ 2\pi^2Te^{-\frac{3}{2}\sqrt{\frac{2n}{n+2}}\phi/M_p}\rho^3\big|_{\xi=\xib}\nonumber\\
 &~~+2\pi^2 \int_\xib^{\xi_{max}} d\xi\, \left[\rho^3\left(\frac{1}{2}\phi'^2+U\right)-3M_p^2(\rho\rho'^2+\rho)\right].
\end{align}
We vary $\phi$ and $\rho$, obtaining four equations (two bulk equations of motion and two surface terms at $\xib$ that must vanish):
\begin{gather}
	\label{eq:EOM_phi} \phi''+\frac{3\rho'}{\rho}\phi' = \frac{\partial U (\phi)}{\partial \phi} , \\
	\label{eq:EOM_rho} \rho'^2 = 1+\frac{\rho^2}{6 M_p^2}\left(\phi'^2-2U(\phi)\right) , \\
	\label{eq:delta_phiprime}\Delta \phi'\big|_{\xib} =-\frac{3}{2}\sqrt{\frac{2n}{n+2}} \frac{T}{M_p} e^{-\frac{3}{2}\sqrt{\frac{2n}{n+2}}\phi/M_p}\big|_{\xib} , \\
	\label{eq:delta_rhoprime}\Delta \rho'\big|_{\xib} =-\frac{\rho T}{2M_p^2}e^{-\frac{3}{2}\sqrt{\frac{2n}{n+2}}\phi/M_p}\big|_{\xib} .
\end{gather}
Here $\Delta\big|_{\xi_b}$ refers to the change in $\phi'$ or $\rho'$ between $\xib^+$  and $\xib^-$ . We assume $\rho$ and $\phi$ are continuous, so that $\Delta U=U_\text{flux}= c e^{ - 3 \sqrt{\frac{2n}{n+2}} \frac{\phi}{M_p}}$ in the approximation described above. 
It is also sometimes useful to have a second-order form of the bulk equation for $\rho$, obtained by differentiating Eq.~(\ref{eq:EOM_rho}),
\begin{align}
	\label{eq:EOM_rho2} \rho'' = -\frac{\rho}{3M_p^2}(\phi'^2+U) .
\end{align}

The boundary conditions needed to construct solutions are determined by the asymptotic form of the BON potential sampled at $\xi=0$; by  equations~(\ref{eq:delta_phiprime}) and~(\ref{eq:delta_rhoprime}) at $\xi=\xib$; and  finally by $ \rho(\xi_{\rm max})= \phi'(\xi_{\rm max}) = 0$ in the case of a false dS minimum, or $\rho\sim\xi$ and $\phi\rightarrow 0$ at large $\xi$ in the case of a Minkowski minimum, for which $\xi_{\rm max}\rightarrow\infty$. We focus on the case of a Minkowski false vacuum, which should give results consistent with results for a de Sitter minimum if the bubble radius is much smaller than the horizon radius~\cite{Draper:2021ujg,Draper:2021qtc}.

\subsection{Approximate Solutions} 

\subsubsection{Initial Conditions}
In the interior region, where the flux is screened, we  assume that the curvature term dominates the potential (we assume $n\geq 2$.) Then we can construct an approximate solution in this regime. Assume  that $\xi\ll R_n$, where $R_n$ is the asymptotic curvature radius of the internal $S^n$. Then 
\begin{align}
\phi \simeq M_p \sqrt{\frac{2n}{n+2}}\log\left(\frac{(n+2)\xi}{2 R_n}\right)\, , \;\;\;\;\;\;\;\;
\rho \simeq \eta R_n\left(\frac{(n+2)\xi}{2 R_n}\right)^\frac{n}{n+2}.
\label{curvsolnappx}
\end{align}
These functions solve the EOM if we drop the ``1" in the equation for $\rho$ and retain only the curvature contribution to $U$. Therefore they only apply at small $\xi$. These leading order $\phi$ and $\rho$ solutions receive corrections from the cc part of the potential that are suppressed by relative factors of $(\xi/R_n)^{4/(n+2)}$. In the large $n \gg 4$ limit these subleading corrections become important earlier as $\xi \rightarrow \mathcal O(R_n)$, but there is always a small $\xi$ regime in which they can be neglected.

In our numerical analysis, $\eta$ will be the shooting parameter that controls the initial condition at $\xi=0$, while $R_n$ and $n$ are fixed inputs. Only those values of $\eta$ that produce the correct asymptotic behavior ($\phi(\xi) \rightarrow \phi_0$ as $\xi \rightarrow \xi_\text{max}$) are counted as BON solutions.
The area radius of the bubble is
\begin{align}
R_3 \equiv \eta R_n,
\end{align}
and the metric in the $\xi \rightarrow 0$ limit is that of $S^3 \times R^n$,
\begin{align}
ds_{4+n}^2 \Big |_{\xi \rightarrow 0} \simeq dr^2 + r^2 \frac{ds_n^2}{R_n^2} + R_3^2 d\Omega_3^2
\label{eq:smallximetric}
\end{align}
for some radial function $r(\xi)\geq 0$. 

\subsubsection{Brane location}
Moving radially outward, eventually we reach the brane at $\xib$. The approximate solution in Eq.~(\ref{curvsolnappx}) is still valid if $\xib/R_n\ll 1$, but this may or may not be the case. Therefore let us first write a general equation that determines $\xi_b$ implicitly, given $R_n$, $n$,  $\eta$, a potential, and a solution for $\xi<\xib$. This amounts to demanding consistency of the $\rho'$ jump condition with the $\rho$ equation of motion, which is first order:
\begin{align}
\frac{\rho}{2M_p^2}Te^{-\frac{3}{2}\sqrt{\frac{2n}{n+2}}\phi/M_p}&=\sqrt{ 1+\frac{\rho^2}{6 M_p^2}\left(\phi'^2-2U|_{Q=0}\right) } \nonumber\\
&-\sqrt{ 1+\frac{\rho^2}{6 M_p^2}\left(\left[\phi'-\frac{3}{2}\sqrt{\frac{2n}{n+2}}\frac{T}{M_p} e^{-\frac{3}{2}\sqrt{\frac{2n}{n+2}}\phi/M_p} \right]^2-2U\right) }.
\label{xibeq}
\end{align}
In this junction condition, $U$ is the total potential $U(\phi)$, including the flux term, and $U|_{Q=0}$ is the potential without the flux term, which is used to solve the equations of motion for $\xi<\xib$. Then $\xib$ is the point where the junction condition is satisfied for a solution of the $\xi<\xib$ equations of motion. 
Depending on the values of $c$ and $T$, the solution to \eqref{xibeq} may reside in the $\xi \ll R_n$ regime where \eqref{curvsolnappx} provides good approximate solutions to $\rho(\xi)$ and $\phi(\xi)$.

Together with the discontinuity equations \eqref{eq:delta_phiprime} and \eqref{eq:delta_rhoprime}, 
\eqref{xibeq} indicates that there is no bubble solution if the brane tension $T$ is taken to zero. That is, with $T \rightarrow 0$, a solution for $\xi_b$ requires $U_{Q=0} = U$. For the flux contribution to the potential this occurs only in the $\phi \rightarrow + \infty$ limit, where the flux potential is negligibly small. 
This is the decompactification limit, not a bubble of nothing decay. For the bubble of nothing, the flux potential is finite at all radii of interest, and so membrane nucleation requires some nonzero tension.

\subsubsection{Beyond the brane}

For $\xi > \xi_b$ the flux is no longer screened, and the flux and curvature parts of the potential both contribute to the solutions for $\phi(\xi)$ and $\rho(\xi)$. 
Analytic approximations for $\rho$ and $\phi$ in this region can be found as series expansions about the curvature-only solutions.
In terms of a convenient parameterization $\lambda(\phi)$,
\begin{align}
\lambda(\phi) \equiv \exp\left( \sqrt{ \frac{2n}{n+2} } \frac{\phi}{M_p} \right),
\end{align}
the solutions for $\lambda$ and $\rho$ can be expressed as
\begin{align}
\lambda(\xi) &= \lambda_{0} + \lambda_c(\xi),
&
\rho(\xi) &= \rho_{0} + \rho_c(\xi),
\\
\lambda_{0} &\approx \left(\frac{n+2}{2} \frac{\xi_b}{R_n} \right)^{\frac{2n}{n+2} } , &
\rho_{0} &\approx R_3 \left(\frac{n+2}{2} \frac{\xi_b}{R_n} \right)^{\frac{n}{n+2} } ,
\end{align}
where $\lambda_c$ and $\rho_c$ are functions that vanish at $\xi = \xi_b$. 
Linearized versions of the equations of motion can be obtained for $\lambda_c \ll \lambda_0$ and $\rho_c \ll \rho_0$,
producing Taylor series solutions for $\lambda(\xi)$ and $\rho(\xi)$,
\begin{align}
\lambda_c(\xi) = \lambda_0 \sum_{m \geq 1} \frac{a_m}{m!} \left( \frac{\xi - \xi_b}{R_n} \right)^m,
&&
\rho_c(\xi) = \rho_0 \sum_{m \geq 1} \frac{c_m}{m!} \left( \frac{\xi - \xi_b}{R_n} \right)^m,
\label{smallseries}
\end{align}
with the first few coefficients satisfying 
\begin{align}
a_2 &= a_1^2 - 3 c_1 a_1 - \frac{6n}{n+2} \frac{ \mathcal C}{\lambda_0^3} + n(n-1) \lambda_0^{- \frac{n+2}{n} } 
, \label{eq:a2}
\\
c_1^2 &= \frac{n+2}{12n} a_1^2 + \frac{R_n^2}{\rho_0^2} - \frac{\mathcal C}{3 \lambda_0^3} + \frac{n(n-1)}{6} \lambda_0^{- \frac{n+2}{n} } .
\label{eq:c12}
\end{align}
Here we introduce a dimensionless $\mathcal C \propto c$ and $\mathcal T \propto T$ as the following combinations of $c$, $T$, $M_p$ and $R_n$:
\begin{align}
\mathcal C &\equiv \frac{R_n^2 c}{M_p^2},
&
\mathcal T &\equiv \frac{R_n T}{M_p^2}.
\end{align}
Eqs.~\ref{eq:a2} and~\ref{eq:c12} provide approximate solutions for $\lambda(\xi)$ and $\rho(\xi)$ in the region close to the brane, $\xi > \xi_b$. Given $\xi_b$, the parameter $a_1$ is determined by the continuity equations \eqref{eq:delta_phiprime} and \eqref{eq:delta_rhoprime}.

In practice we will determine the brane location $\xi_b$ numerically. Although in principle $\xi_b$ could be estimated by combining
\eqref{xibeq} with an analytic approximation for $\rho$ and $\lambda$, we find that the small-$\xi$ limit of \eqref{curvsolnappx} is generally not sufficiently precise to reliably extract $\xi_b$ this way.

\subsubsection{Asymptotics}

Asymptotically far away from the BON, the solution for $\rho(\xi)$ approaches the $\rho \propto \xi$ Minkowski solution, where a generic stabilized potential $U(\phi)$ is approximately
\begin{align}
U(\phi) \approx \frac{1}{2} U_0 \frac{(\phi - \phi_0)^2}{M_p^2} \equiv \frac{1}{2} m_\phi^2 (\phi - \phi_0)^2.
\label{Uasympt}
\end{align} 
In this regime, the solution for $\phi(\xi \gg R_n)$ approaches the false vacuum exponentially quickly,
with $\rho(\xi)$ and $\lambda(\xi)$ taking the form
\begin{align}
\rho(\xi \gg R_n) \simeq \xi + A_m,
&&
\phi - \phi_0 \simeq B_m \frac{K_1(m_\phi \rho) }{m_\phi \rho},
\label{curvsolnasym}
\end{align}
where $A_m$ and $B_m$ are integration constants determined by the details of the $0 \leq \xi \lesssim \mathcal O(R_n)$ solutions.
In the $\xi \gg m_\phi$ limit, $\phi - \phi_0 \propto e^{- m_\phi \xi} (m_\phi \xi)^{-3/2}$. 
As $\phi \rightarrow \phi_0$, the corrections to $\rho' \simeq 1$ become exponentially small, as can be seen from \eqref{eq:EOM_rho}.

\subsection{On-shell action}

To compute the total action of a solution, it is convenient to go back to the form with second-order derivatives of $\rho$:
\begin{align}
S_{tot} &= 2\pi^2 \int_0^\xib d\xi\, \left[\rho^3\left(\frac{1}{2}\phi'^2+U|_{Q=0}\right)+3M_p^2(\rho^2\rho''+\rho\rho'^2-\rho)\right]\nonumber\\
 &~~+ 2\pi^2Te^{-\frac{3}{2}\sqrt{\frac{2n}{n+2}}\phi/M_p}\rho^3\big|_{\xi=\xib}\nonumber\\
 &~~+2\pi^2 \int_\xib^{\xi_{max}} d\xi\, \left[\rho^3\left(\frac{1}{2}\phi'^2+U\right)+3M_p^2(\rho^2\rho''+\rho\rho'^2-\rho)\right]\nonumber\\
 &~~+6\pi^2M_p^2\rho^2\left(\rho'\big|_{\xi=\xib^+}-\rho'\big|_{\xi=\xib^-}\right)\nonumber\\
&~~+\pi^2M_p\sqrt{\frac{2n}{n+2}} \rho^3\phi'\big|_{\xi=0}\nonumber\\
&~~+S_{\rm GHY}
\label{eq:Stot}
\end{align}
where $S_{\rm GHY}$ is the boundary term at infinity given in Eq.~(\ref{eq:sghyformula}). Here the $\Box \phi$ term has been integrated out, removing the modulus boundary term at infinity while introducing a term at $\xi=0$ discussed previously. One advantage of this form is that when we evaluate it on-shell, it simplifies considerably,
\begin{align}
S_{tot} &= -2\pi^2 \int_0^\xib d\xi\,  \rho^3 U|_{Q=0}-2\pi^2 \int_\xib^{\infty} d\xi\, \rho^3 U\nonumber\\
 &~~~~-\pi^2Te^{-\frac{3}{2}\sqrt{\frac{2n}{n+2}}\phi/M_p}\rho^3\big|_{\xi=\xib}\nonumber\\
&~~~~+\pi^2M_p\sqrt{\frac{2n}{n+2}} \rho^3\phi'\big|_{\xi=0}\nonumber\\
&~~~~+6\pi v^2 M_p^2, 
\label{Stotonshell}
\end{align}
where, as discussed above, $v$ is determined by the large-$\xi$ behavior $\rho\sim \xi + const + v/\xi+\dots$. Eq.~(\ref{eq:delta_rhoprime}) has been used to remove the term proportional to $\Delta\rho'$ in the total action, partially cancelling the brane worldvolume term. 

(For numerical purposes the form~(\ref{Stotonshell}) is  more convenient to evaluate than the fully first-order form where $\rho''$ is also integrated by parts. In the latter form there are large-$\xi$ divergences that cancel between the bulk action and the GHY subtraction term $\propto K_0$.)

It is straightforward to check that for $n\geq 2$, the term in the third line of~(\ref{Stotonshell}),  localized at $\xi=0$,  vanishes. Furthermore, if $U$ has a quadratic expansion around the false vacuum, then asymptotically the instanton approaches the false vacuum exponentially fast. Then $v=0$ and the term in the fourth line of~(\ref{Stotonshell}) also vanishes. In these cases we can write the on-shell action compactly as
\begin{align}
S_{tot} &= -2\pi^2 \int_0^\infty d\xi\,  \rho^3 U-\pi^2Te^{-\frac{3}{2}\sqrt{\frac{2n}{n+2}}\phi/M_p}\rho^3\big|_{\xi=\xib}
\label{actionTOT}
\end{align}
with the understanding that the potential is evaluated with zero flux for $\xi<\xib$.

\subsection{Numerical Analysis}
The approximate analytic solutions described above are too qualitative to obtain an accurate result for the tunneling exponent. However, the equations are easily solved numerically with the shooting method. We reformulate the problem by nondimensionalizing the parameters and degrees of freedom, then study several examples numerically.

\subsubsection{Dimensionless parameterization}

In the present formulation, the problem is specified by a set of parameters for the potential, the number of internal dimensions $n$, and the brane tension $T$. The number of free parameters can be reduced by a field redefinition; by applying the constraint that the potential admits a zero-energy minimum; and finally by a convenient length parameter rescaling.

First we define a shifted, dimensionless modulus field $\varphi$ so that the vacuum occurs at $\varphi = 0$:
\begin{align}
\varphi \equiv \frac{\phi - \phi_0}{M_p}.
\end{align} 
Next, we define other dimensionless variables as follows:
\begin{align}
\bar\xi \equiv \frac{\xi \sqrt{U_0} }{M_p},~~~
\bar\rho \equiv \frac{\rho \sqrt{U_0} }{M_p},~~~
u(\varphi) \equiv \frac{U( \phi_0+\varphi M_p)}{U_0},~~~
\overline{ \mathcal T} &\equiv \frac{T\, e^{- \frac{3}{2} \sqrt{ \frac{2n}{n+2} } \frac{\phi_0}{M_p} } }{M_p \sqrt{ U_0} } .
\label{eq:dimlessdef}
\end{align}
Here $U_0$ is arbitrary, but we can think of it as representing a parametric degree of freedom in the potential corresponding to some characteristic energy scale. 

In terms of these variables, the equations of motion take the same form as Eqs.~(\ref{eq:EOM_phi}--\ref{eq:delta_rhoprime}): 
\begin{gather}
	\label{eq:EOM_varphi} \varphi''+\frac{3\bar\rho'}{\bar\rho}\varphi' = \frac{\partial u (\varphi)}{\partial \varphi} , \\
	\label{eq:EOM_rhobar} {\bar\rho}'^2 = 1+\frac{\bar\rho^2}{6 }\left(\varphi'^2-2u(\varphi)\right),\\
	\label{eq:delta_varphiprime}\Delta \varphi'\big|_{\bar\xib} =-\frac{3}{2}\sqrt{\frac{2n}{n+2}} \overline\calT e^{-\frac{3}{2}\sqrt{\frac{2n}{n+2}}\varphi}\big|_{\bar\xib} , \\
	\label{eq:delta_rhobarprime}\Delta {\bar\rho}'\big|_{\bar\xib} =-\frac{\bar\rho \overline\calT}{2}e^{-\frac{3}{2}\sqrt{\frac{2n}{n+2}}\varphi}\big|_{\bar\xib} .
\end{gather}
Likewise the on-shell action \eqref{actionTOT} takes the form
\begin{align}
S_{tot} &= \frac{M_p^4}{U_0} \overline{S},
\label{eq:actiontot}
\\
\overline{S} &\equiv -2\pi^2 \int_0^\infty d\bar\xi\,  \bar\rho^3 u - \pi^2 \overline{\mathcal T} e^{-\frac{3}{2}\sqrt{\frac{2n}{n+2}}\varphi }\bar\rho^3\big|_{\xi=\xib}
.
\label{Stotscaled}
\end{align}
So if $U$ is specified by $m$ parameters, we can absorb one by a field redefinition, eliminate a second by the requirement of having a zero energy minimum, and a third can be absorbed into $U_0$. Then a solution to the system (\ref{eq:EOM_varphi}--\ref{eq:delta_rhobarprime}) for some fixed dimensionless potential $u$ automatically provides a solution to the system (\ref{eq:EOM_phi}--\ref{eq:delta_rhoprime}) for any $U=U_0u$. Furthermore the on-shell action along this family of models is simply obtained by the $1/U_0$ rescaling given in Eq.~(\ref{Stotscaled}). So we really only need to analyze the system for potentials with zero-energy vacua at the origin and a fixed overall scale, corresponding to $m-3$ potential parameters plus the parameter $\overline\calT$ and the integer $n$. 

To proceed we must consider a class of potentials. In our numerical analysis we consider a convenient class where the minimum of the potential is determined to good approximation by the  flux-curvature-CC part of the potential, with $U_\text{other}$ neglected. Then we include the leading effects of a nonzero $U_\text{other}$ by allowing $m_\phi^2$ in Eq.~(\ref{Uasympt}) to float. This alters the curvature at the minimum and the asymptotic behavior of the bubble solutions, but does not move the minimum. 

For such potentials,  the requirement of a zero-energy minimum places a constraint on $(c, R_n, \Lambda_{4+n})$ 
which takes the form
\begin{align}
 c \, (\Lambda_{4+n})^{n-1}= 2^{-n} (n-1)^{2 n-1} R_n^{-2 n}M_p^2.
   \end{align}
Applying this constraint, the Minkowski vacuum is located at
\begin{align}
\phi_0 &= M_p \sqrt{ \frac{n+2}{8n}} \log\left( \frac{(n-1) c }{M_p^2 \Lambda_{4+n} } \right) .
\end{align}
Now we choose
\begin{align}
	U_0 &= c e^{ - 3 \sqrt{\frac{2n}{n+2}} \frac{\phi_0}{M_p}} = c^{-1/2} \left( \frac{M_p^2 \Lambda_{4+n} }{n-1} \right)^{3/2}
	= \left[ \left( \frac{(n-1) M_p^2 }{2 R_n^2} \right)^{3n} \frac{1}{c^{n+2} } \right]^{1/(2n-2) } .
\label{eq:U0}
\end{align}
Replacing $c$ via \eqref{eq:cg}, we find that this $U_0$ is independent of the size $R_n$ of the compact space:
\begin{align}
U_0 &= M_p^4 \left[ (\hat S_n)^6 \left( \frac{2 g_0^2 }{Q^2} \right)^{n+2} \left( \frac{n-1 }{2} \right)^{3n} \right]^{ \frac{1}{2n-2} } ,
&
g &\equiv g_0 M^{\frac{n-4}{2} } , 
\label{eq:U0noR}
\end{align}
where dimensionless $g_0$ sets the strength of the gauge coupling $g$ in units of the fundamental Planck mass $M$,
and where $\hat S_n = S_n/(R_n)^n$ is the area of the compact manifold in units of $R_n$. 
For perturbativity, in the $n < 4$ case we want $g \gtrsim M^{\frac{n-4}{2}}$, i.e.~$g_0 \gtrsim 1$. For $ n> 4$, perturbatively small couplings satisfy $g_0 \lesssim 1$.  

\begin{figure}[t!]
    \centering
    \includegraphics[width=0.55\textwidth]{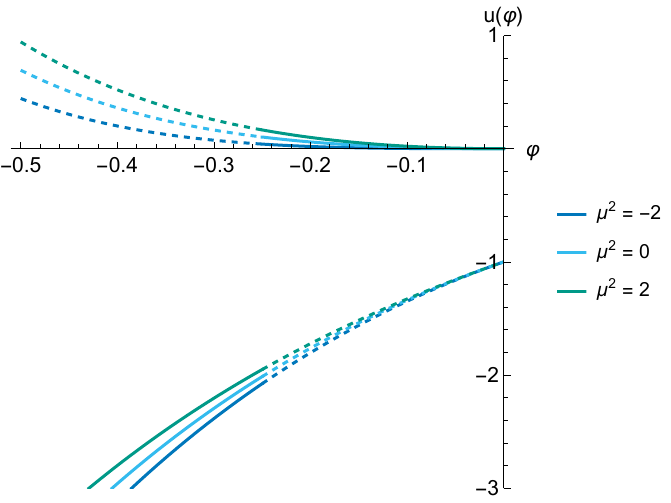}
    \caption{Examples of the  potential  defined in \eqref{eq:dimensionlessU}, with and without the flux term. This figure is drawn for $n = 2$ and various $u_\mathrm{other} = \frac{1}{2} \mu^2 \varphi^2.$ The potential ``seen" by the instanton is shown as a solid line: the trajectory starts at small $\bar \xi$ and large negative $\varphi$ with the flux ``off.'' At some boundary point $\varphi(\bar\xi_b),$ the brane nucleates and the flux is switched on. (This boundary point occurs at different values of $\varphi$ for different solutions; a representative $\varphi_b \approx -0.25$ is used here for illustration.) Stability at the origin bounds $\mu^2\geq -2$.}
    \label{fig:potential}
\end{figure}

Restoring $U_\text{other}$, which can be approximated as $U_\text{other}/U_0\equiv \frac{1}{2}\mu^2 \varphi^2$ for this class of potentials, the full dimensionless $\varphi$ potential takes the form
\begin{align}
u(\varphi) &= \left( e^{ -3 \sqrt{ \frac{2n}{n+2} } \varphi }  - n  e^{ - \sqrt{ \frac{2n+4}{n}} \varphi}  + (n-1)  e^{ - \sqrt{ \frac{2n}{n+2} } \varphi } \right) +\frac{1}{2}\mu^2 \varphi^2.\label{eq:dimensionlessU}
\end{align}
Examples of $u (\varphi)$ are shown in Fig. \ref{fig:potential}.

To summarize, we will study a set of simple models parametrized by a dimensionless brane tension $\overline\calT$, the number of internal dimensions $n$, and the  parameter $\mu^2$ from other contributions to the potential.  We will pick a few values of $n$ and $\mu^2$ and perform more detailed scans over $\overline\calT$. We can also reexpress the superextremality condition in terms of $\overline{\mathcal T} $:
\begin{align}
\overline{\calT}^2 =\frac{T^2}{M_p^2 c} < \frac{2}{3}\left(\frac{n+2}{n-1}\right).
\label{eq:Tmax}
\end{align}

The equations of motion (\ref{eq:EOM_varphi}--\ref{eq:delta_rhobarprime}) for the nondimensionalized parameters $\varphi(\bar\xi)$, $\bar\rho(\bar\xi)$, $u(\varphi)$, and $\overline{\mathcal T}$ do not depend on the values of $U_0$, $Q$, $g$ or $M_p/M$. So, a single numeric solution, e.g.\ for specific values of $n$ and $\mu$, can be translated into physical solutions for any choice of $(Q, g, M_p)$: one simply needs to calculate $U_0$, and to invert \eqref{eq:dimlessdef} to find $\rho(\bar\rho)$, $\xi(\bar\xi)$, etc.
From \eqref{eq:actiontot}, the bounce action depends on $\overline{S}(\bar\rho, \overline{\mathcal T}, u)$, and the ratio $M_p^4/U_0$; 
and from \eqref{eq:U0noR}, it is clear that this combination depends only on the ratio $Q/g$:
\begin{align}
S_\text{tot} = \frac{M_p^4}{U_0} \,\overline{S} 
= \left[ \frac{1}{(\hat S_n)^6} \left( \frac{Q^2}{2 g_0^2 } \right)^{n+2} \left( \frac{2}{n-1 } \right)^{3n} \right]^{ \frac{1}{2n-2} } \, \overline{S} ,
\label{eq:StotQ}
\end{align}
where for the $n$-sphere $\hat S_n = 2 \pi^{\frac{n+1}{2} } /\Gamma(\frac{n+1}{2} )$.

\subsubsection{Examples}

We turn now to numerical investigation. We  integrate the equations of motion for $\varphi$ and $\bar \rho$ and use the shooting method to determine the value of the parameter $\eta$ defined in (\ref{curvsolnappx}) 
 (recall that this sets the bubble size). The initial conditions are given by (\ref{curvsolnappx}) at some small $\xi_0$, with modifications to account for the shift $\phi_0$ relating $\phi$ and $\varphi$ and the switch to the dimensionless $\bar\xi$:
\begin{align}
    \varphi \left(\bar\xi\right) \simeq \sqrt{\frac{2n}{n+2}} \log \left(\frac{n+2}{\sqrt{2\left(n-1\right)}} \bar \xi\right) \, , \;\;\;\;\;\;\;\;
    \bar \rho \left(\bar\xi\right) \simeq \eta \left(\frac{n + 2}{2} \bar \xi \right)^{\frac{n}{n+2}}
\label{curvsolnappxDimensionless}
\end{align}
at small $\bar\xi$. 
For each trial $\eta$,  integration proceeds until the jump conditions (\ref{eq:delta_varphiprime}--\ref{eq:delta_rhobarprime}) are satisfied, at which point the brane is inserted and integration continues with the modified potential. For a given value of $\overline{\calT},$ a solution to the system requires not only the fields $\bar \rho$ and $\varphi$ but also the correct brane location $\bar \xi_b$.

For illustration we show results in the cases $n = 2$ and $n = 6$, and for a variety of potentials $u_\mathrm{other}$ as defined in Eq.~(\ref{eq:dimensionlessU}). In order to justify the treatment of $u_\mathrm{other}$ as a perturbation, we restrict $| \mu^2 | \le | d^2  u / d \varphi^2 |_{\mu^2 = 0}.$ 
\begin{figure}[t!]
    \centering
    \includegraphics[width=0.42\textwidth]{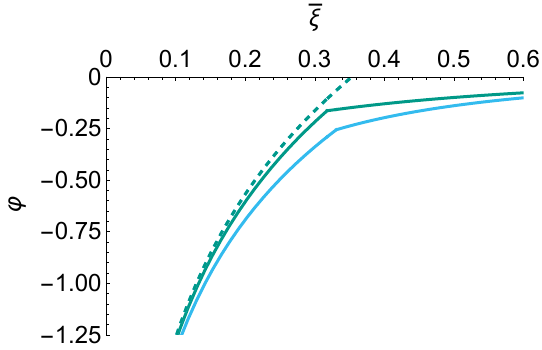}~~~~
    \includegraphics[width=0.5\textwidth]{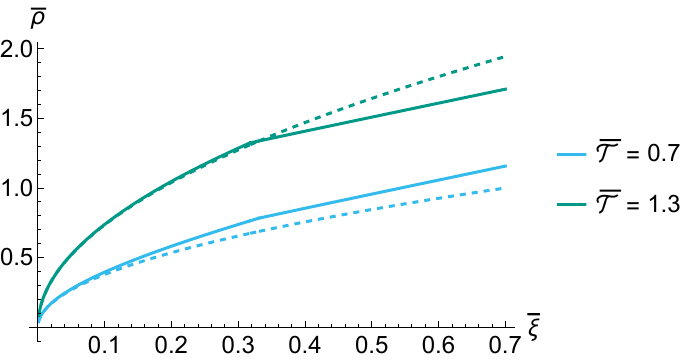}
    \caption{The behavior of $\varphi$ and $\bar \rho$ for two values of $\mathcal{\bar T},$ with $n = 2$ and $\mu^2 = 0.$ A discontinuity occurs in the derivative at the brane boundary $\bar \xi_b,$ just above $\bar\xi\sim0.3$ in both cases shown, after which $\varphi$ exponentially approaches 0 and $\bar \rho$ increases linearly. The dotted lines represent the small-$\bar \xi$ approximations given in \ref{curvsolnappxDimensionless}.}
    \label{fig:functionsApprox}
\end{figure}
Examples of solutions for $\bar \rho$ and $\varphi$ are shown in Fig.~\ref{fig:functionsApprox}. Confirming the approximations in \eqref{curvsolnappx} and \eqref{curvsolnasym}, $\varphi$ is logarithmic inside of the brane and $\bar \rho$ follows an approximate power law, while outside of the brane $\bar \rho$ becomes linear and $\varphi$ exponentially approaches zero.

\begin{figure}[t]
    \centering
    \begin{subfigure}[b]{0.45\textwidth}
    \includegraphics[width=\textwidth]{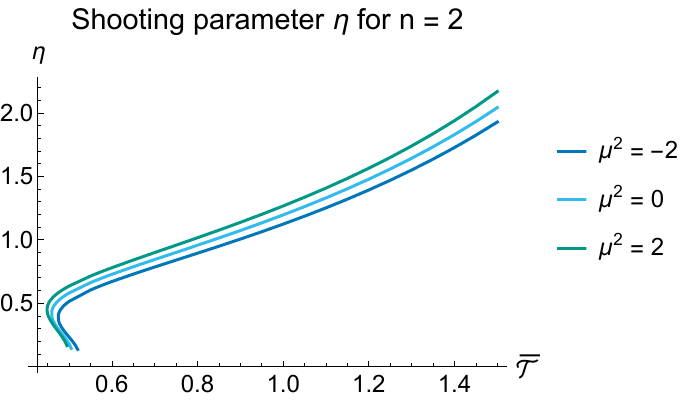}
    \caption{}
    \label{fig:TvEtaA}
    \end{subfigure}
    \begin{subfigure}[b]{0.45\textwidth}
    \includegraphics[width=\textwidth]{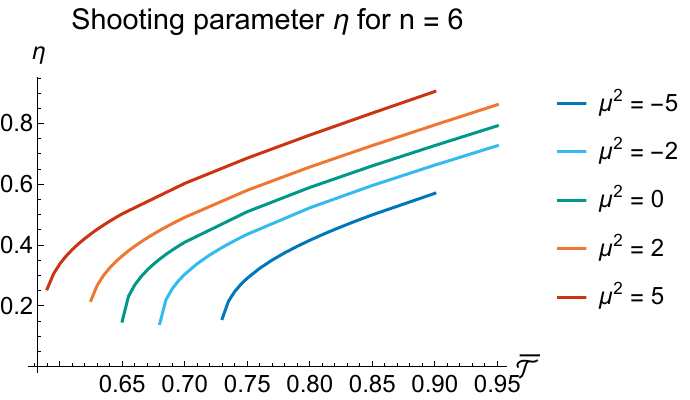}
    \caption{}
    \label{fig:TvEtaB}
    \end{subfigure}
    \caption{The value of the shooting parameter $\eta$ as defined in (\ref{curvsolnappxDimensionless}). It decreases rapidly close to $\mathcal{\overline T}_{min}$, but is approximately linear for larger values of $\mathcal{\overline T}.$}
    \label{fig:TvEta}
\end{figure}

\begin{figure}[t!]
    \centering
    \begin{subfigure}[b]{0.45\textwidth}
    \includegraphics[width=\textwidth]{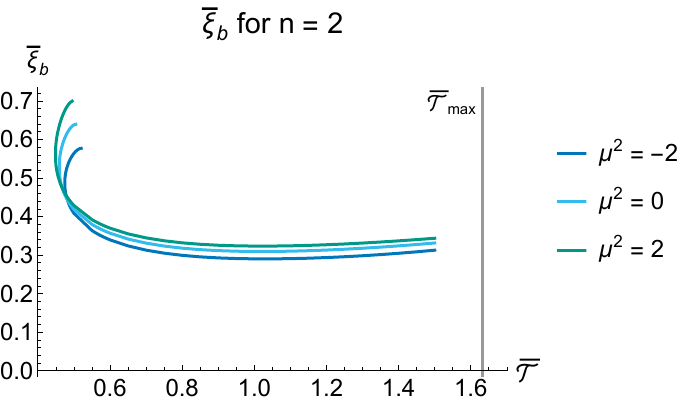}
    \caption{\label{fig:TvXiA}}
    \end{subfigure}
    \begin{subfigure}[b]{0.45\textwidth}
    \includegraphics[width=\textwidth]{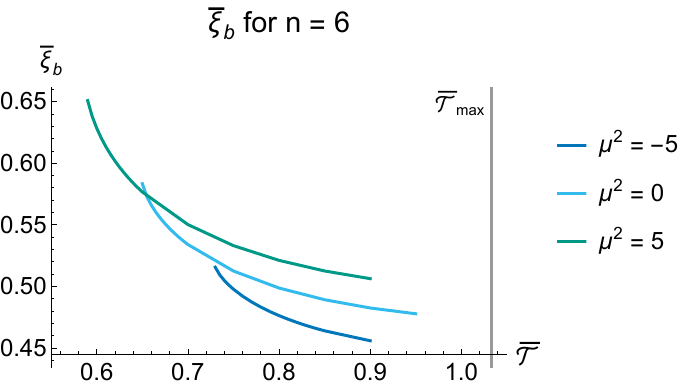}
    \caption{\label{fig:TvXiB}}
    \end{subfigure}
    \caption{The membrane nucleation radius $\bar \xi_b$ as a function of the tension. Left:  $n = 2$. Here $\bar \xi_b$ reaches a shallow minimum near $\overline \calT \sim 1$, then increases steeply for  $\overline \calT\sim 0.5$ where it converges with an upper branch of solutions.  Right: $n=6$. Here no minimum $\overline\calT$ was found before the shooting code reached limits of numerical precision, but the trend is suggestive of similar behavior. The extremality bound on $\overline \calT$ from Eq.~\eqref{eq:Tmax} is marked on each graph.}
    \label{fig:XiGraph}
\end{figure}

For certain values of $\overline{\calT},$ we found that there are (at least) two branches of solutions corresponding to two different shooting parameters/bubble sizes $\eta$, and two different brane positions $\bar \xi_b$. This is evident at low $\overline{\calT}$ in Figs.~\ref{fig:TvEta} and ~\ref{fig:TvXiA}. A similar multi-branched phenomenon was found in the flux-free cases studied in~\cite{Draper:2021qtc}. The branches converge at a minimum value of $\overline{\calT},$ below which we found no viable solutions. In the lower branch, smaller $\bar \xi_b$ corresponds to a large negative $\varphi$ at the brane. Therefore the slope discontinuities and the potential energy difference $\Delta u|_{\bar \xi_b}$ across the brane,  which depend on $\exp (-\varphi),$ are relatively large. This branch also has generally larger values of the shooting parameter $\eta.$ Kinematically, on this branch the field $\varphi$ begins at $\xi = 0$ with a large energy, which it loses almost all at once in order to nucleate a brane with a large jump in the potential energy. In comparison, the upper branch has relatively large $\bar \xi_b$ and hence a smaller magnitude of $\varphi\left(\bar \xi_b\right)$; the system spends more of its energy inside of the boundary and loses only a small amount to nucleating the brane.

\begin{figure}[t!]
    \centering
    \begin{subfigure}[b]{0.45\textwidth}
    \includegraphics[width=\textwidth]{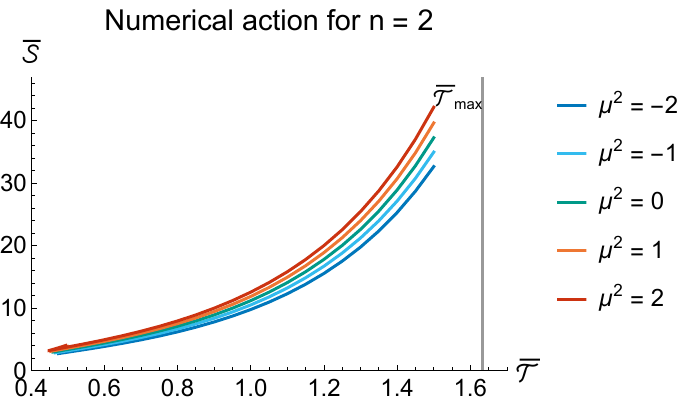}
    \caption{\label{fig:SGraphA}}
    \end{subfigure}
    \begin{subfigure}[b]{0.45\textwidth}
    \includegraphics[width=\textwidth]{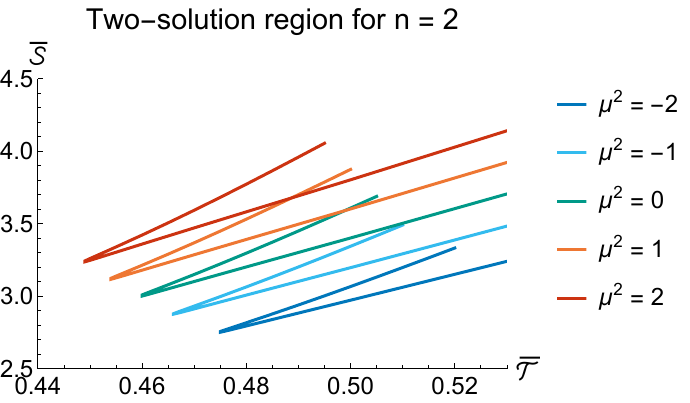}
    \caption{\label{fig:SGraphB}}
    \end{subfigure}
    \begin{subfigure}[b]{0.45\textwidth}
    \includegraphics[width=\textwidth]{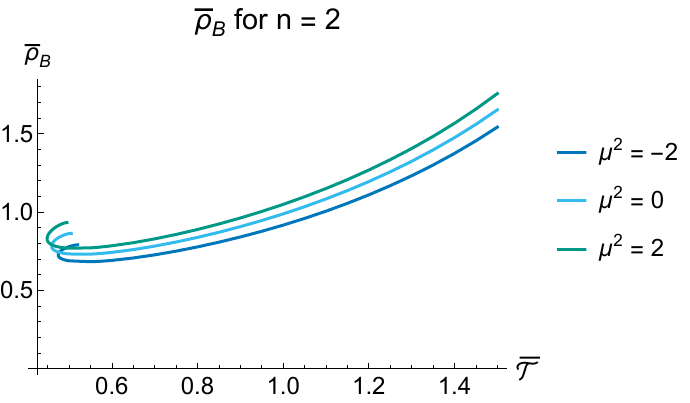}
    \caption{\label{fig:SGraphC}}
    \end{subfigure}
    \begin{subfigure}[b]{0.45\textwidth}
    \includegraphics[width=\textwidth]{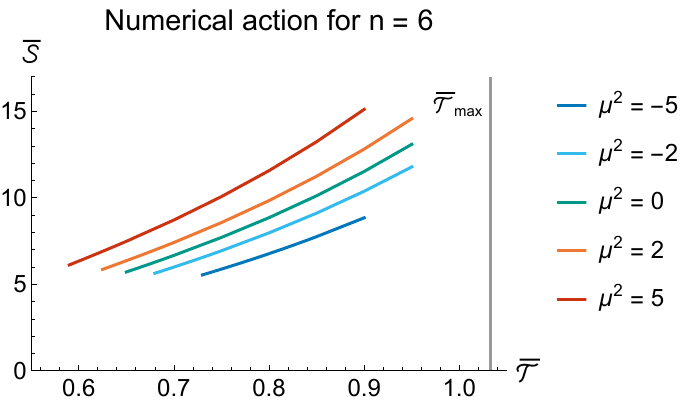}
    \caption{\label{fig:SGraphD}}
    \end{subfigure}
    \caption{Values of the scaled action $\mathcal{\bar S}$ as a function of brane tension for various mass parameters $\mu$.  Top left: $n=2$. Results for the upper branch of $\bar\xi_b$ solutions are difficult to see on this plot, so in the top right we show a zoom-in of the corner where these solutions appear. The two branches converge at some $\overline \calT_\mathrm{min}$, below which no solutions were found. Lower plot:  $n = 6$. We did not find a second solution branch for the $n = 6$ case. The vertical line at $\overline{\mathcal T}_\text{max}$ indicates the extremality bound \eqref{eq:Tmax}. \ref{fig:SGraphC} shows $\bar \rho_B$ in the $n = 2$ case, which is useful in understanding the behavior of the action.}
    \label{fig:SGraph}
\end{figure}

From the numerical profiles we can compute the scaled action in Eq.~\eqref{Stotscaled}. Results are shown in Fig.~\ref{fig:SGraph}. For $n=2$, the lower  branch of $\bar \xi_b$ solutions  generically has the lower value of the action, indicating that it is the dominant branch responsible for tunneling. On the upper branch, we found viable solutions only for restricted values of $\overline{\calT}$; this is due to numerical difficulties as the space of viable solutions becomes small in this region and the shooting method becomes unreliable. Since this branch is not associated with the decay process (and likely has multiple fluctuation modes of negative eigenvalue) we do not attempt to explore it with more sophisticated methods. On the lower branch, we find that solutions exist up to near the extremal tension $\overline{\calT}_{\mathrm{max}}$ calculated from \eqref{eq:Tmax}. At this point, similarly, the space of solutions is small and the shooting method becomes intractable.

 The on-shell action grows faster than linearly with $\overline{\mathcal T},$ although only the boundary term has an explicit, linear, $\overline{\mathcal T}$ dependence. This behavior can be understood by noting that, away from the two-solution region, $\eta$ is approximately linear in $\overline{\mathcal T}$, cf. Fig.~\ref{fig:TvEta}. In turn, for a given value of $\bar \xi < \bar \xi_b$, $\bar \rho$ scales close to linearly with $\eta$ as a consequence of the initial condition. $\varphi$ is somewhat less sensitive since the initial condition is independent of $\eta$. The behavior of (\ref{Stotscaled}) with $\overline{\mathcal T}$ is then controlled by explicit and implicit dependence in $\bar\rho$.

 As a result, although tunneling can always proceed through black branes, the on-shell action is typically smaller and the tunneling rate faster if the theory contains fundamental charged brane excitations (superextremal branes). However, this effect is limited by the lower bound on $\calT$, which is not typically far from the extremal limit.

\begin{figure}[t!]
    \centering
    \begin{subfigure}[b]{0.45\textwidth}
    \includegraphics[width=\textwidth]{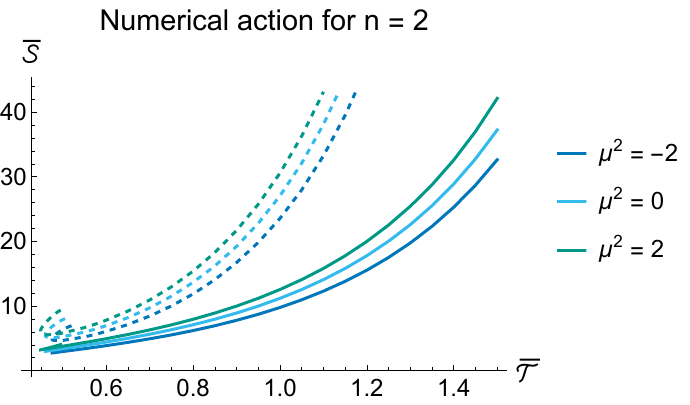}
    \caption{\label{fig:TotalvBoundaryA}}
    \end{subfigure}
    \begin{subfigure}[b]{0.45\textwidth}
    \includegraphics[width=\textwidth]{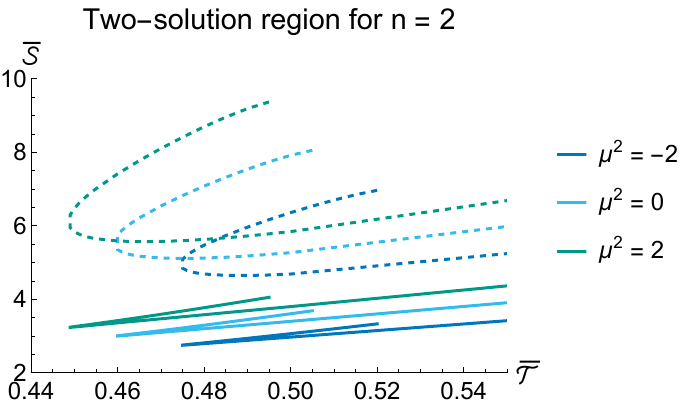}
    \caption{\label{fig:TotalvBoundaryB}}
    \end{subfigure}
    \begin{subfigure}[b]{0.45\textwidth}
    \includegraphics[width=\textwidth]{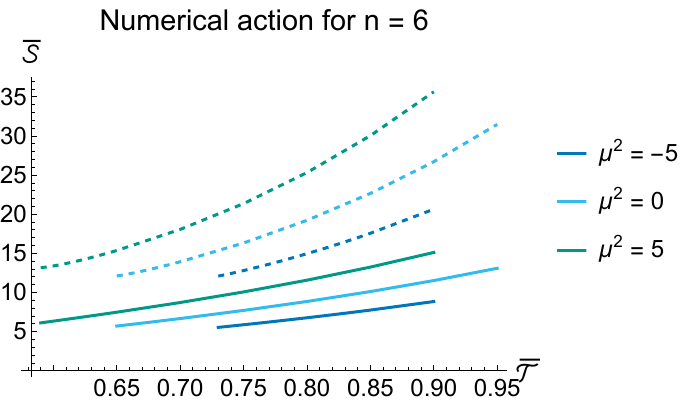}
    \caption{\label{fig:TotalvBoundaryC}}
    \end{subfigure}
    \caption{A comparison of the total on-shell action (solid lines) with the brane contribution (dotted lines). Because the action contains negative terms, the brane contribution is larger than the total action. The brane contribution, however, does not show a sharp cusp at low $\mathcal{\overline T}$.}
    \label{fig:TotalvBoundary}
\end{figure}

It is also interesting to compare the  brane contribution to the action (the second term in (\ref{eq:Stot})) to the total. This is shown in Fig.~\ref{fig:TotalvBoundary}. In fact, because the action also contains negative terms, the brane contribution is actually somewhat larger than the overall action.  (Note that the brane and bulk contributions to the action are smooth in the vicinity of $\overline{\mathcal T}_\text{min}$, even though the total action is cuspy.)

\subsection{Summary}

 If the BON process is not otherwise forbidden, bubbles form with a nucleation rate per volume proportional to $\Gamma \sim v^4 \exp(-S_E)$, which for smaller radii $R$ can exceed one bubble per Hubble volume. 
 The effects of a BON would be cataclysmic for any observers in its forward light cone. Our existence and ability to measure the age of the universe sets a bound on the smallness of any compact extra dimensions which admit such decays. 

 Stabilizing the moduli that set the sizes of the extra dimensions typically requires several ingredients, including classical and quantum effects. If fluxes are involved, charged objects must also nucleate to screen the flux before the bubble can form.
 Our analysis includes charged branes in the spectrum of states, so that the bubble rate can be estimated using the spherically symmetric Coleman-De~Luccia ansatz, with the charged brane inserted at some self-consistently-determined radial distance from the bubble center.

We improve on previous analyses by providing a general solution algorithm for internal $n$-spheres and by studying the parametric dependence on the brane tension. 
In the examples we studied, it is exponentially faster to nucleate a fundamental, superextremal brane than an extremal black brane. However, typically there is a minimum brane tension $T_\text{min}$ below which it is not possible to solve the junction condition. 
The lower limit on $T$ corresponds to the merging of two branches of Schwinger-BON solutions in the $n=2$ case, where the branch with larger total action positions the charged brane farther away from the bubble center. Although the brane location varies rapidly as a function of $T \gtrsim T_\text{min}$, a cancellation between the bulk and brane terms causes the total action for the two branches to remain relatively close in magnitude.

A constraint on  extra dimensional model parameters can be obtained from the lifetime of the universe, by requiring that the bubble nucleation rate per unit volume $\Gamma$ should be small compared to the Hubble size, $\Gamma / H_0^4 \lesssim 1$. Estimating $\Gamma \sim R_n^{-4} \exp( - S)$, the corresponding limit on the bubble action is $S \gtrsim 560 - 4\log(M_p R_n)$, or
\begin{align}
 \overline{S} \gtrsim \frac{U_0}{M_p^4} \left( 560 - 4\log(M_pR_n) \right) ,
\end{align}
for $U_0(c, R_n)$ given in \eqref{eq:U0}.

In the $n=2$ case, the bubble action is quite sensitive to $T$, changing by an order of magnitude between $T \approx T_\text{min}$ and $T \approx 0.9 T_\text{max}$. 
Taking points near $T_\text{min}$ and $T_\text{max}$ as an example, with $\overline{S} (\mathcal T = 0.5) \approx 3$ and $\overline{S} (\mathcal T = 1.25) \approx 52$ for the $\mu = 0$ model, the universe lifetime sets a lower limit on $Q$: from \eqref{eq:StotQ}, the $n=2$ bounds are
\begin{align}
\overline{\mathcal T} \simeq 0.5: ~~~ |Q| \gtrsim 20\, g_0 ,
&&
\overline{\mathcal T} \simeq 1.25: ~~~ |Q| \gtrsim 10\, g_0 ,
\end{align}
where we neglect the logarithmic term  for the sake of simplicity.

For $n=6$, the scaling $S_\text{tot} \propto Q^{8/5}$ is less steep, and the lower bound on $Q$ occurs at correspondingly larger values. Taking the $\mathcal T = 0.65$ and $\mathcal T = 0.95$ points from the $\mu =0$ line in Fig.~\ref{fig:TotalvBoundary}, we find for $n=6$:
\begin{align}
\overline{\mathcal T} \simeq 0.65: ~~~ |Q| \gtrsim 150\, g_0 ,
&&
\overline{\mathcal T} \simeq 0.95: ~~~ |Q| \gtrsim 96\, g_0 .
\end{align}
Although $Q$ here is integer-valued, $g_0$ parameterizes the strength of the gauge coupling, and can take any perturbatively small value.

In each case, relatively strong couplings $g \sim M^\frac{n-4}{2}$ (i.e.~$g_0 \sim \mathcal O(1)$) require the compact space to be stabilized with multiple units of flux: $Q \sim 10$ for $n=2$, and $Q \sim 10^2$ for $n=6$. 
For $n > 4$, weaker couplings correspond to smaller values of $g_0$; so, our lower bound on $Q$ is saturated by $|Q| \geq 1$ once $g_0 \lesssim 10^{-2}$ in the $n=6$ case.

For $n=2$, on the other hand, perturbatively small couplings $g$ require $g_0 \gtrsim 1$ (see \eqref{eq:U0noR}), i.e.~$1/g \lesssim M^{\frac{4-n}{2}}$. So, a small gauge coupling $1/g_0 \sim 10^{-2}$ requires large charges $|Q| \gtrsim 10^{3}$ in order to stabilize the compact dimension against BON decays mediated by branes of tension $T \gtrsim T_\text{min}$.

\section{Wilson Lines and Incompatible Fermions}

We turn now to a different class of obstruction and its resolution. Consider 5D Kaluza-Klein theory and a background, flat U(1) gauge field with a nonzero Wilson line around the $S^1$, $A_{5}=const$.  In the bubble geometry a nonzero $A_5$ is incompatible with $dA=0$ by Stokes' theorem. 

This situation arises in the low energy description of Kaluza-Klein theory with a massive fermion and non-antiperiodic boundary conditions. The fermion boundary conditions obstruct the BON, but we need a way of communicating this obstruction to the gravitational EFT where the fermion has been integrated out. To do so we introduce a nondynamical U(1) gauge field $A$ coupled to fermion number and perform an improper gauge transformation $\psi\rightarrow e^{-i A_{5,0}\phi^5}\psi,~A_5\rightarrow A_5-A_{5,0}$. For suitable choice of $A_{5,0}$ the fermion becomes antiperiodic and the topological obstruction has been transferred to the background gauge field in the sense described above. At this point the fermion can be integrated out.

Alternatively, we can consider massless fermions and leave them in the theory, but continue to shuffle the topological obstruction into a background gauge field. 

At this point it is clear what is needed to restore the BON instability: the gauge field must be made dynamical, and in the instanton solution we require that $A_5$ relaxes to zero at the bubble wall. The Euclidean action contains the Maxwell term
$
\frac{1}{4g^2}\int \sqrt{g} d^5x F_{ab}F^{ab}
$
 and if $A_5$ obtains a radial profile, then it contributes directly to the on-shell action and back-reacts on the geometry. 

This phenomenon was first described in~\cite{Blanco-Pillado:2016xvf}, in the  context of 4D supergravity on $\RtSo$. There are various issues with $R$-symmetry in 4D supergravity~\cite{KS1,KS2,Distler:2010zg,Seiberg:2010qd}, and in the model of~\cite{Blanco-Pillado:2016xvf} the gauged $U(1)_R$ is Higgsed at the Planck scale. Here we will consider simpler examples on $\RfSo$. In the case of an ordinary Dirac fermion coupled to Einstein-Maxwell theory, and in 5D supergravity with a gauged, unbroken $U(1)_R$, the relevant bubble of nothing solution can be written down analytically and corresponds to the Euclidean continuation of  a 5D Reissner-Nordstrom black hole.\footnote{
Other topological solitons with shrinking $S^1$ include the black strings and topological stars of~\cite{Bah:2020ogh,Bah:2020pdz,Bah:2021irr} and the Lorentzian solitons in $\mathcal N = 2$ and $\mathcal N=8$ supergravity of~\cite{Anabalon:2021tua,Anabalon:2022aig}.
}

Before proceeding we must comment on the effects of moduli potentials. First, we will ignore the stabilizing potential for the radial modulus. This is only a good approximation of the scale of the potential is somewhat small compared to the Einstein-Hilbert term, e.g. $V \lesssim  M_p^2/R^2$ for characteristic bubble curvature scale $\sim $ KK scale $1/R$, but we make it for simplicity. There is also a ``holonomy potential" for  $A_5$  generated by loops of charged particles circumnavigating the extra dimension. Therefore, whether the vacuum structure is such that the modification described above is actually required is a dynamical question. If there are  light species with $mR\ll 1$ the potential is dominated by them, and the leading contribution is~(see, e.g.,~\cite{Arkani_Hamed_2007})
\begin{align}
V(A_{5}) = \sum_{\rm \tiny DOF\, i}(-1)^{F_i+1} \frac{3}{64\pi^6 R^4}{\rm Re\, 
Li}_5(e^{i(\theta_i-2\pi q_iA_{5})}).
\end{align}
Here $\theta_i=0$ ($\theta_i=\pi$) for naively periodic (antiperiodic) degree of freedom and $F_i=0\ (1)$ for bosons (fermions). We are interested in cases where the fermions are naively antiperiodic and the bosons are naively periodic, so that there is no  obstruction at the bubble wall if $A_5\rightarrow 0$ there. Furthermore we are particularly interested in cases where the $A_5$ vacuum satisfies $2\pi q_i\langle A_5\rangle  \neq 0 \,({\rm mod}\, 2\pi) $ for at least some $i$, so that the gauge field vev cannot be removed by an improper transformation without changing at least one of these naive periodicities. (In this case a bubble of nothing would exist, but it would be an ordinary one, with no gauge field profile.) In fact, for massless matter the best we can do is to establish metastable minima with this property. (For example, one $q=1$ fermion and $N$ $q=2$ bosons leads to a global minimum at $A_5=0$ and a local minimum at $A_5=1/2$, which become degenerate for large $N$.) Alternatively, and perhaps more simply, we may take all matter to be massive compared to the KK scale $1/R$. In that case the $A_5$ effective potential from charged loops is exponentially suppressed $\sim e^{-mR}$, and it may receive competing contributions from  other  UV sources. The result is model-dependent and we can simply assume it has a relevant minimum away from the origin. Going forward we will make this assumption, and neglect the effects of $V(A_{5})$ on the instanton, since they are subleading in the $\hbar$ expansion.

\subsection{Bubble and decay rate}
The relevant solution to the Einstein equation, $G_{\mu\nu}=\kappa T_{\mu\nu}$, and the Maxwell equation, $\partial_\nu (\sqrt{g} F^{\mu\nu})=0$,  is  5D Euclidean Reissner-Nordstrom,
\begin{align}
ds^2 &= V^{-1}dr^2+r^2 d\Omega_3^2 + R_0^2 V d\phi_5^2\nonumber\\
A_5 &=A_{5,0}\left(1-\frac{b_+^2}{r^2}\right)\nonumber\\
V&= \left( 1 - \frac{b_+^2}{r^2} \right)\left( 1 + \frac{c}{R_0} \frac{b_+^2}{r^2} \right) 
\nonumber\\
b_\pm&=R_0\pm c\nonumber\\
c&=\frac{2A_{5,0}^2\kappa}{3 g^2 R_0}
\label{bonsol}
\end{align}
Here $\kappa = 8\pi G_N$ is the 5D Newton constant,  $g$ is the 5D gauge coupling (mass dimension $-1/2$), 
 $R_0$ is the proper radius of the KK circle at infinity, $A_{5,0}\sim A_{5,0}+\mathbb{Z}$ is the vacuum value of the $\phi_5$ component of the gauge field at infinity,  and $\phi_5\sim \phi_5+2\pi$.\footnote{The corresponding RN solution has charge $Q=\frac{2A_{5,0}b_+^2}{g^2 R_0}$, which can also be thought of as the instanton charge under the zero form gauge field (axion) in the dimensionally reduced theory. It is generally large in the semiclassical regime, permitting nearly arbitrary $A_{5,0}$ to be compatible with charge quantization. We thank Matt Reece for a discussion of this point.} The bubble radius is
\begin{align}
R_3 = b_+
\end{align}
where  $V$ vanishes. In the limit $A_{5,0}\rightarrow 0$ it reduces to Witten's bubble, Euclidean Schwarzschild. The solution admits the same growing-bubble Lorentzian continuation, $\theta\rightarrow i\theta+\pi/2$, where $\theta$ is a polar angle of the $S^3$.

The leading order decay rate is $e^{-S_E}$, where the Euclidean action $S_E$ is a sum of bulk gauge, bulk Einstein-Hilbert, and boundary GHY terms. The total is
\begin{align}
S_E &= \frac{2\pi^3 b_+^3}{\kappa}\nonumber\\
&=\frac{\pi^2\left(R_0+\frac{2A_{5,0}^2\kappa_4}{3g_4^2R_0}\right)^3}{R_0\kappa_4}\nonumber\\
&\approx\frac{\pi^2R_0^2}{\kappa_4}+\frac{2\pi^2 A_{5,0}^2}{g_4^2}.
\end{align}
The final line is valid for $1/g_4^2 \ll R_0^2 /\kappa_4$. 

This solution demonstrates that when bubbles of nothing are topologically obstructed by spin structure, the obstructions can sometimes be removed by coupling to dynamical gauge fields.

 \subsection{Dirac spectrum}
 The discussion in this section is somewhat tangential. In the subclass of theories with massless fermions, it is interesting to investigate the spectrum of the Dirac operator on the bubble background~(\ref{bonsol}). If there is an isolated zero mode then the functional determinant will vanish.\footnote{Continuum zero modes are generally integrable and do not lead to a vanishing functional determinant.}
 There is no particular reason the Dirac operator should have an isolated zero mode, but it is interesting that there is a straightforward procedure to check it explicitly without computing the entire determinant, exploiting the fact 
 that the instanton is conformal to a product manifold~\cite{Cho:2007zi}.\footnote{The Ricci scalar is positive definite, so if there was no gauge field, we could  appeal to the Schr\"odinger-Lichnerowicz equation to immediately conclude that the spectrum is positive definite. The presence of the gauge field complicates the problem somewhat.}
 
The  Dirac operator is
\begin{align}
\slashed{D}&= \gamma^\mu D_\mu= \gamma^\mu(\partial_\mu - \frac{i}{4}\omega^{ab}_\mu\sigma_{ab}-iA_\mu)
\end{align}
where $\gamma^\mu = \gamma^a e^\mu_a$, $\omega$ is the spin connection, $\omega^{ab}=\frac{i}{2}[\gamma^a,\gamma^b]$, and the vielbein can be set to $e^a_\mu=\sqrt{g_{a\mu}}$.
 Following~\cite{Cho:2007zi} we take
\begin{align}
\tilde g_{\mu\nu} &= \Omega^2 g_{\mu\nu},~~~~~
\tilde \psi = \Omega^{-2} \psi,~~~~~
\tilde A  =  A,~~~~~
\slashed{D}\psi = \Omega^3 \tilde{\slashed{D}}\tilde \psi,~~~~~
\Omega=1/r.
\end{align}
Then
\begin{align}
\widetilde{ds^2} &= \frac{dr^2}{V r^2}+ \frac{R_0^2 V}{r^2} d\phi_5^2+d\Omega_3^2 
\label{conformal}
\end{align}
and the $r-\phi_5$ direction has separated from the 3-sphere. If the Dirac operator has a zero mode on~(\ref{conformal}) then it will also have one on~(\ref{bonsol}), so we can examine the spectrum on either manifold; we choose to treat the conformal transformation as a field redefinition and continue to study the  eigenvalue problem on the original spacetime. In the new variables it can be written in tensor product form,
\begin{align}
i[(\tilde\gamma^r \tilde D_r+\tilde\gamma^{\phi_5}\tilde D_{\phi_5})\otimes 1 + \tilde\gamma^5\otimes(\tilde\gamma^\sigma\tilde D_\sigma)_{S^3}]\tilde\psi = \lambda r \tilde\psi,
\end{align}
where the $r$ on the right-hand side results from the conformal transformation.  
We  expand $\tilde\psi$ in eigenspinors on the 3-sphere,
$\tilde\psi = \sum_\ell\left(\phi_\ell^{+}\chi_{\ell}^{+}+\phi_\ell^{-}\chi_{\ell}^{-}\right)$, 
where $\phi^{\pm}=\phi^{(\pm)}(r,\phi_5)$ are radial modes and
\begin{align}
(\tilde\gamma^\sigma\tilde D_\sigma)_{S^3}\chi^{\pm}_\ell = \pm i\left(\ell +\frac{3}{2}\right)\chi^{\pm}_\ell .
\end{align}
 Henceforth we drop the tildes to simplify the notation. The $\phi$ equation becomes
 \begin{align}
i\left[(\gamma^r  D_r+\gamma^{\phi_5} D_{\phi_5}) \pm i\left(\ell +\frac{3}{2}\right)\gamma^5\right]\phi_{\ell}^{\pm}
=\lambda r \phi_{\ell}^{\pm},
\label{eq:2dirac}
\end{align}
a  2-dimensional Dirac equation with $\gamma^5$ interaction and  potential.
We can choose the  Dirac matrices to be $\gamma^r=r\sqrt{V}\sigma^2$ and $\gamma^{\phi_5}=\frac{r}{R_0\sqrt{V}}\sigma^3$, and we set $\gamma^5=-\sigma^1$, which anticommutes with the others and squares to the identity. Note that if $\chi_\ell$ is a solution to~(\ref{eq:2dirac}) with ``$+$" sign and parameter $\lambda$, then $\gamma^5\chi_\ell$ is a solution to ~(\ref{eq:2dirac}) with ``$-$" sign and parameter $-\lambda$.  We define
\begin{align}
\phi^{\pm} = e^{i  \phi_5/2}  \sqrt{r} V^{-1/4}
\bigg(
\begin{array}{c}
F_\ell^{\pm}(r)\\
G_\ell^{\pm}(r)
\end{array}
\bigg),
\label{phiFG}
\end{align}
corresponding to a low-lying KK mode consistent with antiperiodic boundary conditions. For $\langle A_5\rangle=1/2$,  the effective asymptotic boundary condition around the $S^1$ is periodic and the mass of the KK mode vanishes in the vacuum. In terms of $F$ and $G$, Eq.~(\ref{eq:2dirac}) becomes
\begin{align}
\bigg(
\begin{array}{cc}
\frac{A_5-1/2}{R_0\sqrt{V}}&\sqrt{V}\partial_r \pm\frac{1}{r}\left(\ell+\frac{3}{2}\right)\\
-\sqrt{V}\partial_r \pm\frac{1}{r}\left(\ell+\frac{3}{2}\right)&-\frac{A_5-1/2}{R_0\sqrt{V}}
\end{array}
\bigg)
\bigg(
\begin{array}{c}
F_\ell^{\pm}(r)\\
G_\ell^{\pm}(r)
\end{array}
\bigg)
=\lambda \bigg(
\begin{array}{c}
F_\ell^{\pm}(r)\\
G_\ell^{\pm}(r)
\end{array}
\bigg) .
\label{fulldirac}
\end{align}
These equations can be separated and put in normal form. The result is two Schr\"odinger-like equations for $F$ and $G$, in which $\lambda$ plays the role of a potential parameter rather than the energy, and the ``energy" is zero.  The complete expressions for the potentials in the Schr\"odinger equations are somewhat complicated and for numerical purposes it is easier to work with Eq.~(\ref{fulldirac})  directly. However the separated equations are convenient for studying the asymptotic behaviors which we briefly examine first. 
We  fix $\langle A_5\rangle=1/2$ as described above.

\begin{figure}[t!]
\centering
\includegraphics[width=0.85\textwidth]{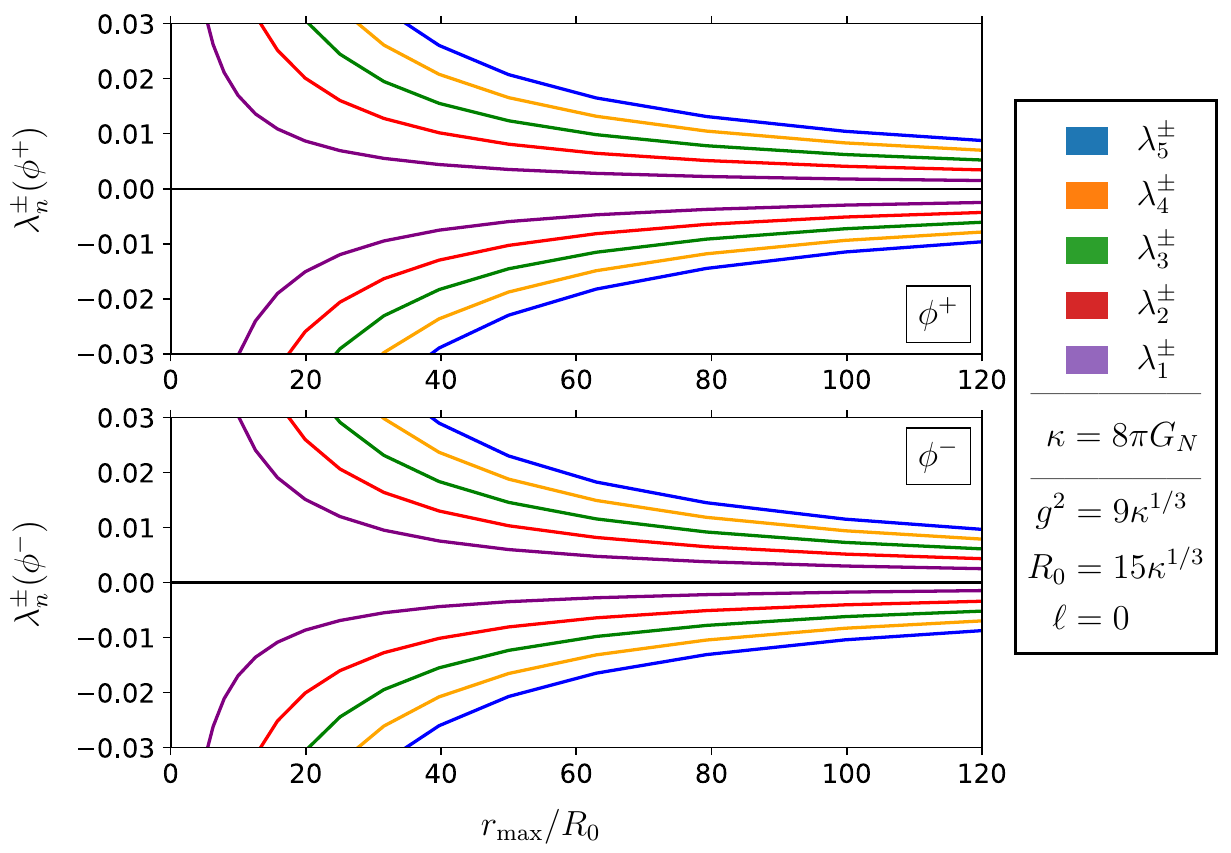}
\caption{Here we plot the values of the first five positive and negative $\ell = 0$ eigenvalues as a function of $r_\text{max}$, for $F^+$ and $G^+$ (top) and $F^-$ and $G^-$ (bottom). In this example we take $g^2 = 9 \kappa^{1/3}$ and $R_0 = 15 \kappa^{1/3}$.
In the $\phi^+$ case, the positive $\lambda_n$ are smaller in magnitude than the corresponding negative eigenvalues; for $\phi^-$, it is the negative eigenvalues that have the smaller magnitudes. In the $r_\text{max} \gg R_0$ limit, $\lambda$ scales as $\lambda \propto 1/r_\text{max}$.
}
\label{fig:lambda}
\end{figure}

Near the bubble wall, where $V=A_5=0$, we set $r = R_3+x$ and expand in small $x$. The leading form of the separated equations is
\begin{align}
-\partial_x^2 F_\ell^{\pm} -\frac{1}{x} \partial_x F_\ell^{\pm}+\frac{1}{16x^2}F_\ell^{\pm}=0
\end{align}
and the same for $G_\ell^{\pm} $, independent of $\lambda$ and the model parameters. In  normal form these become Schr\"odinger equations with an attractive $3/16x^2$ potential. 
The regular solutions behave as $G_\ell^{\pm}= F_\ell^{\pm}\sim x^{1/4}$ at small $x$; this cancels the $x^{-1/4}$ singularity in the factor of $V^{-1/4}$ in Eq.~(\ref{phiFG}). 
At large $r$, the leading behavior of the separated equations is
\begin{align}
-\partial_r^2 F_\ell^{\pm} + \left(\frac{(2\ell+3)(2\ell+3\mp 2)}{4r^2}\right)F_\ell^{\pm} = \lambda^2 F_\ell^{\pm}
\end{align}
and the same equation is satisfied by $G_\ell^{(\mp)}$. We see there is a repulsive $1/r^2$ potential even in the case $\ell=0$. At some intermediate $r$ the potential will reach a  positive maximum.

To go further it is simplest to solve~(\ref{fulldirac}) numerically. This can be done for arbitrary $g$, $R_0$, and $\ell$. We put the problem on an interval $r\in [R_3,R_{\rm max}]$ for some large $R_{\rm max}$. The boundary conditions are determined by requiring that $i\slashed{D}$ is self-adjoint with respect to the inner product $\int d^5x \sqrt{g} \psi_1^\dagger\psi_2$. It is sufficient to require that $i\hat{{\slashed{D}}}\equiv i(\tilde\gamma^r \tilde D_r+\tilde\gamma^{\phi_5}\tilde D_{\phi_5})$ is self-adjoint with respect to the inner product $\int \frac{dr}{r^2}  \phi_1^\dagger\phi_2$, where the weight function $1/r^2$ arises from the invariant measure on the $r-\phi_5$ factor in~(\ref{conformal}). $i\hat{{\slashed{D}}}$ is already formally self-adjoint, and becomes truly self-adjoint if we adopt the boundary conditions that $F$ and $G$ are regular at $r=R_3$ and $F(r_{\rm max})=G(_{\rm max})$. 
We use a simple point-and-shoot method to solve~(\ref{fulldirac}), imposing Dirichlet boundary conditions and searching for solutions with no $F=G$ nodes to identify the $n=\pm1$ eigenvalues. Higher modes in the spectrum ($|n|>1$) were found by searching for solutions with one or more nodes within the bulk region $0 < r < R_\text{max}$. 

\begin{figure}
\centering
\includegraphics[width=\textwidth]{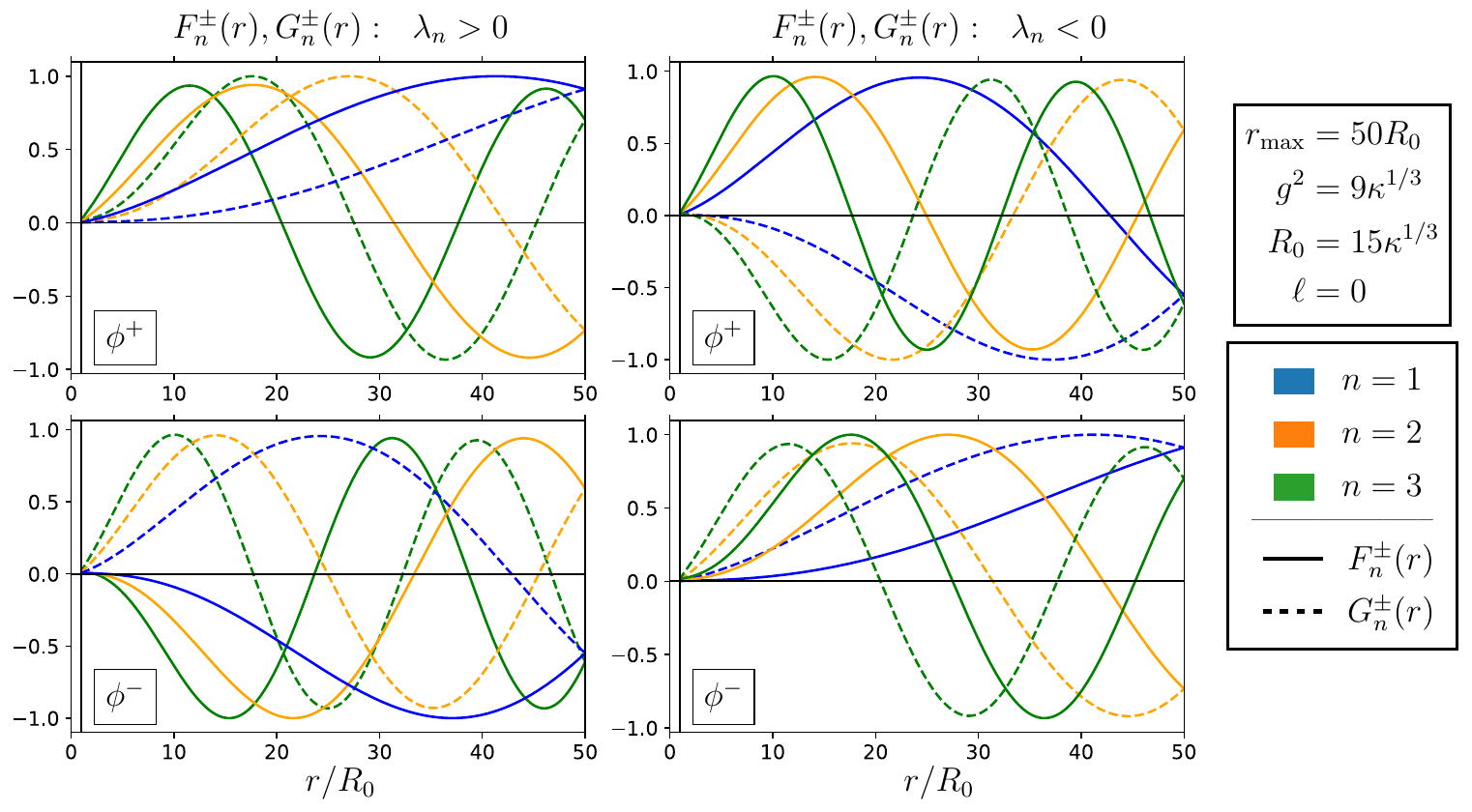}
\caption{In this example we show the first three $\ell = 0$ eigenfunctions $F^\pm_n(r)$ and $G_n^\pm(r)$, $n=1, 2, 3$, for a specific value of $r_\text{max} = 50 R_0$. The number of nodes increases with the value of $n$. Each panel shows $F^\pm_n$ as a solid line, and $G^\pm_n$ as dashed. The top and bottom rows show $\phi^+$ and $\phi^-$, while the left and right columns correspond to positive and negative eigenvalues, respectively. Note that $F^\pm_n(r) =  G^\mp_{-n} (r)$, where $\pm n$ indicates the $n$th positive or negative eigenvalue. }
\label{fig:FGr}
\end{figure}

\begin{figure}
\centering
\includegraphics[width=\textwidth]{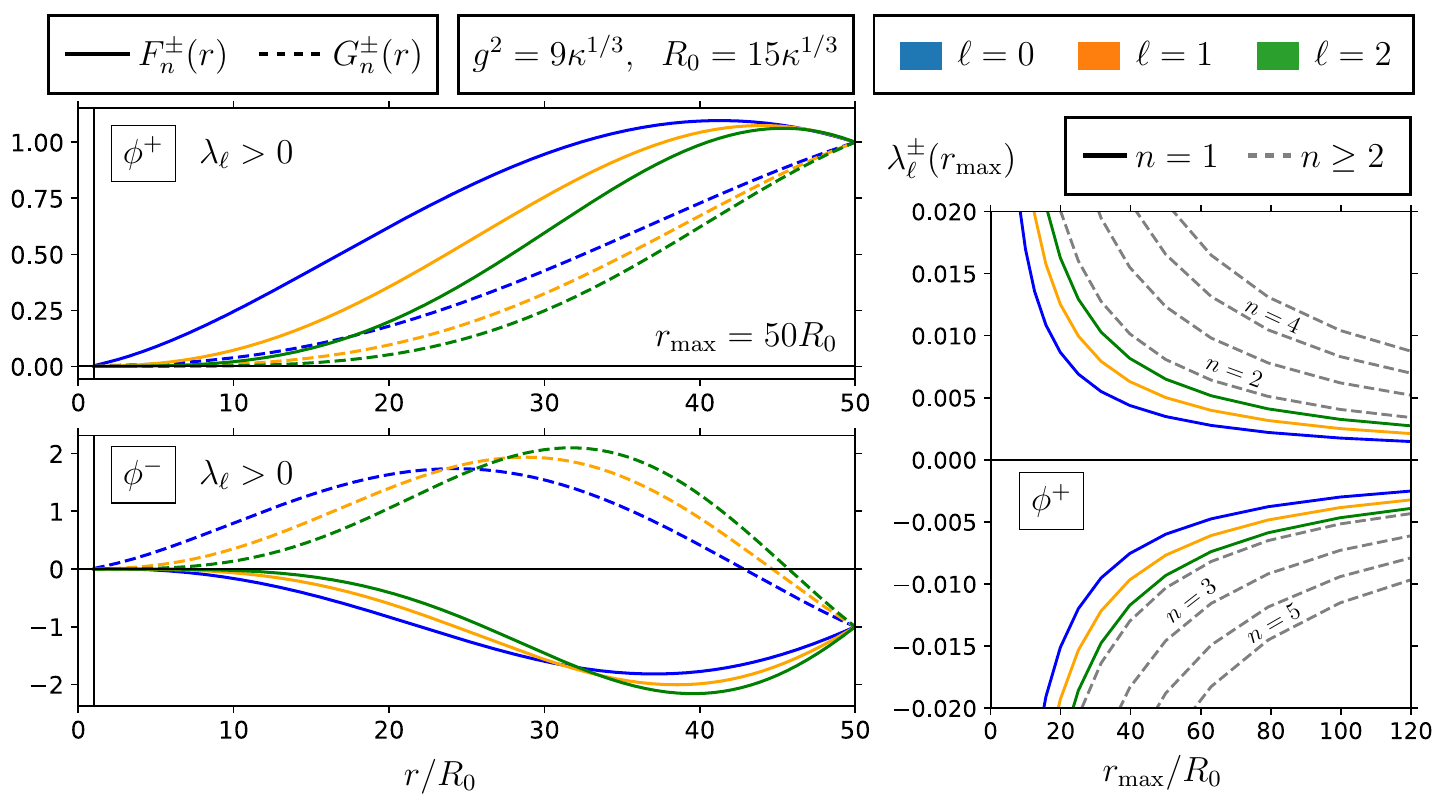}
\caption{\textbf{Left:} We plot the $F^\pm$ and $G^\pm$ eigenfunctions for $n=1$ and $\ell = 0, 1, 2$, for an example with $r_\text{max} = 50 R_0$. The upper and lower panels show $\phi^+$ and $\phi^-$ respectively, with positive $\lambda_\ell > 0$ in both cases. As demonstrated in Figure~\ref{fig:FGr}, the $\lambda < 0$ solutions can be obtained from the positive eigenvalue solutions by the transformation $F^\pm \leftrightarrow G^\pm$, $\lambda_n \leftrightarrow \pm \lambda_{-n}$. In each case we have normalized the eigenfunctions to $F^\pm(r_\text{max}) = G^\pm(r_\text{max}) = \pm 1$ for visual clarity.
\textbf{Right:} Here we show the eigenvalues $\lambda_{n, \ell}$ as a function of $r_\text{max}$ for the $F^+$, $G^+$ type solutions. The solid lines show $\ell = 0, 1, 2$ for $n=1$; these solutions have the smallest magnitudes $|\lambda_{n,\ell}|$. The gray dashed lines show the $n = 2, 3, 4, 5$ eigenvalues with $\ell =0$. As in Figure~\ref{fig:lambda}, the eigenvalues for $\phi^-$ can be obtained from the values above by applying the transformation $\lambda_n \leftrightarrow \pm \lambda_{-n}$.  }
\label{fig:ell}
\end{figure}

Example results are shown in Figs.~\ref{fig:lambda},\,\ref{fig:FGr},\,\ref{fig:ell}. In Fig.~\ref{fig:lambda} we show the low-lying spectrum as a function of the radial cutoff $r_{\rm max}$ for fixed choices of model parameters $R_0$ and $g$. The eigenvalues all converge to zero together like $1/{r_{\rm max}}$ for large cutoff, forming the continuum of unbound modes with continuous radial wavevector in the infinite volume limit. The corresponding eigenfunctions are shown in Fig.~\ref{fig:FGr} for fixed large $r_{\rm max}$. As expected, no isolated zero mode is observed.  Fig.~\ref{fig:ell}  shows results for other values of $\ell$.

\subsection{Supergravity}
It appears that the mechanism discussed above can operate in supergravity, but only in rather exotic circumstances.\footnote{
In $\mathcal N = 2$ supergravity there are Lorentzian soliton solutions with a shrinking $S^1$~\cite{Anabalon:2021tua}. 
}  
On-shell, minimal 5D supergravity contains a gravitational field $e^a_m$, a gravitino $\psi_m$, and a graviphoton $A_m$. The main issue is the gravitino, which must be coupled to a gauge field in order for the Reissner-Nordstrom-type relaxation of the spin structure to operate. The gravitino is a doublet under the global $SU(2)_R$, so we need to gauge part of this symmetry. A method of gauging a $U(1)$ subgroup of the $R$ symmetry was discussed by Zucker~\cite{Zucker:1999fn}. The on-shell Lagrangian is essentially Einstein-Maxwell theory with two massive spin-3/2 fields coupled to the gauge field, a Chern-Simons term that will not be relevant for our purposes, and a negative cosmological constant. Let us explicitly break SUSY by adding soft terms to set the gravitino mass and c.c.\ to zero. The important feature is that the gravitino kinetic term contains the gauge coupling
\begin{align}
-\frac{i}{2}\bar\psi_p \gamma^{p5n}(\partial_5 -iA_5 \sigma^2)\psi_n.
\end{align}
If the $\psi_n$ are periodic, we can transform $\psi_n\rightarrow e^{i\phi^5\sigma^2/2} \psi_n$. This makes the fermions antiperiodic around the $S^1$ while turning on a constant background field $A_5=1/2$. In other words, it is exactly the situation we found above. So the Euclidean Reissner-Nordstrom solution again describes a vacuum decay process in this softly-broken gauged supergravity theory.

\section{Discussion}
Instantons that mediate topology change may be singular if the vacuum contains background gauge fields. As in nongravitational examples, dynamical sources and gauge fields can regulate the singularities, leading to well-defined, finite-action saddle points of the path integral. Here we briefly summarize the findings of this paper and some directions for future work.

Bubble of nothing tunneling in the presence of flux-stabilized internal manifolds requires co-nucleation with a charged membrane. Our work contributes to the literature on this topic by providing general, bottom-up tools for constructing approximate instanton geometries and a careful treatment of boundary terms in the on-shell action. We also study the dependence on the brane tension, finding that while fundamental, superextremal branes facilitate exponentially faster transitions, there is a lower bound on the tension for which such solutions exist, and it is not parametrically below the extremal limit. However, our analysis relies on an approximation in which the flux is fully screened inside the brane nucleation radius, as though the brane were smeared across the internal manifold instead of localized at a point, and it is not completely clear whether a more physically accurate treatment would preserve these properties. The flux term in the modulus potential is nonsingular inside the brane  radius, so it should rapidly become subdominant to the curvature term, but a more careful study of the uncertainties associated with the full-screening approximation is warranted.

Wilson operators for background gauge fields provide an alternative description of the topological obstruction caused by non-antiperiodic fermions on $S^1$, and provide a convenient EFT description of the obstruction below the fermion mass. If the gauge field is  made dynamical, it can relax the periodicity at the bubble wall so that nucleation can proceed. We studied the simplest instanton in this class, which arises in ordinary Kaluza-Klein theory, and it would be interesting to extend the analysis to cases in which Wilson operators for higher-form gauge fields obstruct bubble of nothing decays with higher-dimensional internal manifolds.

\vskip 1cm
\noindent
{\bf Acknowledgements:}  
We thank Isabel Garcia~Garcia for early collaboration on this project.
This work was supported in part by the U.S.
Department of Energy, Office of Science, Office of High Energy Physics under award number DE-SC0015655.
The work of B.L.\ was supported in part by the U.S. Department of Energy under grant number DE-SC0011640.

\bibliography{bon_refs}{}

\bibliographystyle{utphys}

\end{document}